# State-of-the-art SPH solver DualSPHysics: from fluid dynamics to multiphysics problems

J.M. Domínguez[1*], G. Fourtakas[2], C. Altomare[3,4], R.B. Canelas[5], A. Tafuni[6], O. García-Feal[1], I. Martínez-Estévez[1], A. Mokos[7], R. Vacondio[8], A.J.C. Crespo[1], B.D. Rogers[2], P.K. Stansby[2], M. Gómez-Gesteira[1]

[1]EPHYSLAB Environmental Physics Laboratory, CIM-UVIGO, Universidade de Vigo, Spain

[2]Department of Mechanical, Aerospace and Civil Engineering (MACE), University of Manchester, United Kingdom

[3]Maritime Engineering Laboratory, Universitat Politècnica de Catalunya – BarcelonaTech, Barcelona, Spain

[4]Ghent University, Department of Civil Engineering, Technologiepark 60, 9052 Gent, Belgium

[5]Bentley Systems, Lisbon, Portugal

[6]School of Applied Engineering and Technology, New Jersey Institute of Technology, Newark, US

[7]Saint-Venant Hydraulics Laboratory, Ecole des Ponts ParisTech, Paris, France

[8]Department of Engineering and Architecture, University of Parma, Parma, Italy

*Corresponding author. E-mail address: jmdominguez@uvigo.es, ORCID: 0000-0002-2586-5081

## Abstract

DualSPHysics is a weakly compressible smoothed particle hydrodynamics (SPH) Navier-Stokes solver initially conceived to deal with coastal engineering problems, especially those related to wave impact with coastal structures. Since the first release back in 2011, DualSPHysics has shown to be robust and accurate for simulating extreme wave events along with a continuous improvement in efficiency thanks to the exploitation of hardware such as graphics processing units (GPUs) for scientific computing or the coupling with wave propagating models such as SWASH and OceanWave3D. Numerous additional functionalities have also been included in the DualSPHysics package over the last few years which allow the simulation of fluid-driven objects. The use of the discrete element method (DEM) has allowed the solver to simulate the interaction among different bodies (sliding rocks, for example), which provides a unique tool to analyse debris flows. In addition, the recent coupling with other solvers like Project Chrono or MoorDyn has been a milestone in the development of the solver. Project Chrono allows the simulation of articulated structures with joints, hinges, sliders and springs and MoorDyn allows simulating moored structures. Both functionalities make DualSPHysics one of the meshless model world leaders in the simulation of offshore energy harvesting devices. Lately, the present state of maturity of the solver goes beyond single phase simulations, allowing multi-phase simulations with gas-liquid and a combination of Newtonian and non-Newtonian models expanding further the capabilities and range of applications for the DualSPHysics solver. These advances and functionalities make DualSPHysics a state-of-the-art meshless solver with emphasis on free-surface flow modelling.

# 1. Introduction

The meshless method smoothed particle hydrodynamics (SPH) has made great progress over the past two decades with many contributions to the method continually appearing. Modern SPH schemes enable the simulation of violent free-surface flows with improved accuracy, while new physical processes can be included with relative ease accompanied by a significant reduction of the computational time achieved through hardware acceleration.

The development of SPH enables the simulation of different physical phenomena, including violent hydrodynamics of coastal/offshore waves impacting structures, galaxy and planetary formation and evolution, flow in process industries characterised by multiple phases and mixing, large deformations of solids and structures, modelling damage and failure, including opening and propagation of cracks, fluid-structure interaction, and exploitation in computer graphics and computer games (see [1–4] for recent examples). The most relevant advantages over Eulerian methods [5] are the exact conservation of mass and momentum and the meshless properties. Despite these advances, some key challenges still remain, and are encapsulated in the SPHERIC Grand Challenges (https://spheric-sph.org/grand-challenges).

With the large numbers of neighbouring particle interactions and the time step restricted by the Courant condition for weakly-compressible flow, SPH is computationally demanding and there have been multiple efforts within the broad SPH community to accelerate the simulations. Massive parallelisation is the traditional approach where machines with tens of thousands of processing cores are used in a combination of distributed and shared memory frameworks [6–8]. In the past 10 years, the exploitation of hardware such as graphics processing units (GPUs) for scientific computing has gained traction. SPH is particularly suitable to graphics processing unit (GPU) acceleration thanks to its vector lists of particles and their interactions. Multiple GPU-based SPH codes now exist [9–11].

The open-source code DualSPHysics is being developed precisely to address these issues. The DualSPHysics project is an open-source SPH package with a growing user community, including annual international Users Workshops and 52,000 downloads in total with up to 10,000 downloads for each sub-release. In a highly collaborative approach, the Universidade de Vigo, University of Manchester, University of Lisbon, Università di Parma, Flanders Hydraulics Research, Universitat Politécnica de Catalunya, and New Jersey Institute of Technology have developed DualSPHysics, which represents the state-of-the-art for exploiting the computational power of modern GPUs and enables simulation of real engineering problems using a standard desktop PC. The open-source code is freely available at https://dual.sphysics.org/ and is released under GNU Lesser General Public License (LGPL).

DualSPHysics' name derives from being able to run on both central processing unit (CPU) using the shared memory OpenMP approach, or on a GPU. The main thrust of the DualSPHysics development has been to exploit the computational acceleration using the compute unified device architecture (CUDA) programming framework on Nvidia GPUs. One of the guiding philosophies for this dual functionality is that new features are developed first in C++ language for the CPU version, and once validated they are more easily ported to a GPU using CUDA. This can facilitate new developers to implement their own customized solutions without learning CUDA, which is less widespread than C++. This approach allows engineers to focus on developing new functionalities, while computer scientists can work on ensuring computational efficiency of the GPU version. The code has highly optimised CUDA kernels for a single GPU, it uses hierarchical templates and optimized cell-linked neighbour lists [12] to achieve maximum code flexibility and processing speed [13].

DualSPHysics comes with dedicated pre- and post-processing tools. In particular, the GenCase application is used to generate all the necessary inputs. In the XML input files constants, geometries, and parameters are specified. Recently, a graphical user interface (GUI) plugin for the open-source software FreeCAD, namely DesignSPHysics (https://design.sphysics.org/), has made the workflow for the generation of new test cases more straightforward and accessible. Different post-processing tools are available inside the DualSPHysics package, either to visualise the simulation using any scientific visualisation software such as Paraview or to extract computed quantities of interest (vorticity, surface elevation, forces, etc.).

The DualSPHysics code has evolved continuously since its first release in 2011. The most recent release (version 5.0, July 2020) includes many relevant novelties: (i) Coupling with the discrete element method and Project Chrono (https://projectchrono.org/); (ii) Coupling with MoorDyn (http://www.matt-hall.ca/moordyn.html); (iii) a multi-phase solver (soil-water and gas-liquid); (iv) specialised functionality for second-order wave generation and active wave absorption, plus relaxation zones; (v) coupling with wave propagation models; and (vi) inlet/outlet flow conditions. Some limitations remain related to the SPHERIC Grand Challenges (GC3: Adaptivity, GC4: Coupling, GC5: Applicability to industry), which are the focus of ongoing research.

The aim of this paper is to present the novel developments of the DualSPHysics software, emphasising the process of validation that is undertaken before software releases. The paper is organised as follows: Section 2 describes the core SPH formulation. Section 3 is dedicated to the description of different features: wave generation, fluid-structure interaction, inflow-outflow, and multi-phase formulations. Section 4 presents one of the key developments of coupling with other models, including wave propagation models, DEM, the physics engine Project Chrono and the library MoorDyn. Section 5 presents the new code implementation, which has changed significantly over the lifespan of the code. In Section 6 validations and applications are presented. Finally, the current challenges are briefly discussed in Section 7 in the context of ever-changing hardware, user demands and ongoing research.

## 2. SPH formulation

### 2.1. Interpolants and kernel functions

The integral (continuous) representation of SPH in a domain $\Omega$ for a smooth function $f(\mathbf{r}), \mathbf{r} \in \mathbb{R}^d$ can be written as the convolution a kernel function $W$ and $f$

$$\langle f \rangle(\mathbf{r}) \coloneqq \int_\Omega f(\mathbf{r}')W(\mathbf{r}-\mathbf{r}',h)d\mathbf{r}', \tag{1}$$

where the $\langle \cdot \rangle$ symbol denotes an SPH interpolant and will be dropped henceforth for simplicity, $\mathbf{r}' \in \Omega$ is a temporary position variable and $W(\mathbf{r}, h)$ is a positive function with the general form of,

$$W(\mathbf{r},h) \coloneqq \frac{1}{h^d}\omega(|q|), \tag{2}$$

with $h > 0$ and $q = |\mathbf{r}|/h$. The function $\omega: \mathbb{R} \to \mathbb{R}$ is a smooth non-negative function such that,

$$\int_\Omega \omega(q)d\mathbf{r} = 1, \tag{3}$$

with compact support (see Figure 1),

$$\omega(q) = 0 \ for \ |\mathbf{r}| \geq kh, \ k \in \mathbb{R}^+. \tag{4}$$

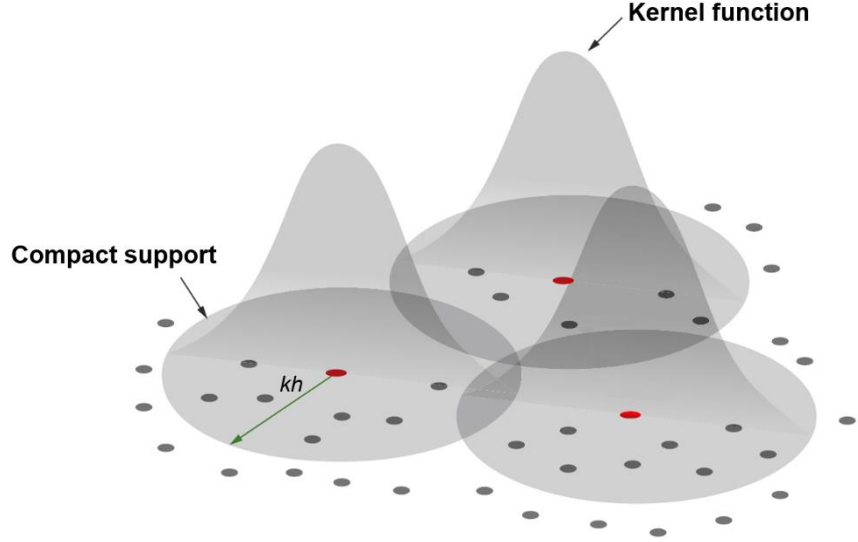

Figure 1. Shape and compact support of overlapping kernel functions.

The discrete convolution of $W$ and $f$, defined herein as the SPH discrete approximation, is obtained by

$$f(\mathbf{r}_a) := \sum_{b \in \mathrm{P}} f(\mathbf{r}_b) W(\mathbf{r}_a - \mathbf{r}_b, h) \Delta r_b^d, \tag{5}$$

where the subscripts $a, b \in \mathrm{P} = \{1, \cdots, N\}$ denote the interpolating and neighbouring discrete particles, while $\Delta r_b^d$ is the associated volume $V$ of the $b$-th particle in P. The set P contains all particles within the domain i.e. fluid F, boundary B and floating objects K as shown in Figure 2 with $\mathrm{P} = \mathrm{F} \cup \mathrm{B}$ and $\mathrm{K} \subset \mathrm{B}$.

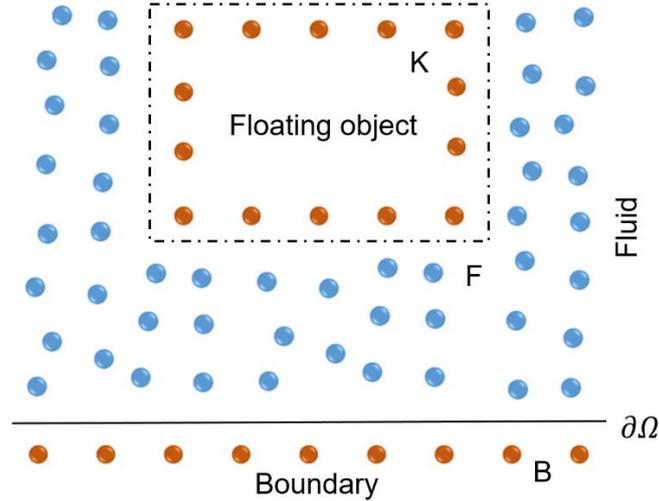

Figure 2. Sketch of different sets of particles $a \in \mathrm{P}$.

For a more complete description of SPH, the reader is referred to [14, 15] and a recent review in [5]. Similar integral and discrete expressions can be formulated for a gradient operator for a given function $f$ by,

$$\frac{\partial f(\mathbf{r})}{\partial \mathbf{r}} = \int_\Omega f(\mathbf{r}') \frac{\partial W}{\partial \mathbf{r}}(\mathbf{r} - \mathbf{r}', h) d\mathbf{r}', \tag{6}$$

by employing the convolution identity,

$$\frac{\partial f}{\partial \mathbf{r}} * W = f * \frac{\partial W}{\partial \mathbf{r}}. \qquad (7)$$

The discrete gradient takes the following form,

$$\frac{\partial}{\partial \mathbf{r}} f(\mathbf{r}_a) := \sum_{b \in P} f(\mathbf{r}_b) \frac{\partial}{\partial \mathbf{r}} W(\mathbf{r}_a - \mathbf{r}_b, h) V_b, \quad a = 1, \cdots, N. \qquad (8)$$

Several smoothing kernels have been proposed in the literature as summarised in [16]. The smoothing kernel or simply kernel takes the form of Eq. (2) and satisfies Eq. (3) and (4), with $W: \mathbb{R}^d \to \mathbb{R}$. Moreover, it is positively defined within the support domain $\{\forall \mathbf{r}' \in \Omega; W(\mathbf{r} - \mathbf{r}', h) > 0\}$, it is a monotonic decreasing sufficiently smooth function defined over the interval $kh$ and,

$$\lim_{h \to 0} W(\mathbf{r}, h) = \delta(\mathbf{r}), \qquad (9)$$

where $\delta$ is the Dirac delta function [17, 18]. Several kernel functions have been implemented in DualSPHysics.

The widely used third order B-splines kernel (cubic spline) reads [18]

$$W(\mathbf{r}, h) = a_D \begin{cases} 1 - \frac{3}{2}q^2 + \frac{3}{4}q^3 & 0 \le q \le 1 \\ \frac{1}{4}(2 - q)^3 & 1 < q \le 2 \\ 0 & otherwise \end{cases}, \qquad (10)$$

where $\alpha_D$ is $10/7\pi h^2$ and $1/\pi h^3$ in 2-D and 3-D respectively.

The Wendland function [19], namely the $C^2$ kernel which exhibits a positive kernel Fourier transform [16, 20] and better pairing instability characteristics. The Wendland $C^2$ reads

$$W(\mathbf{r}, h) = a_D \left(1 - \frac{q}{2}\right)^4 (2q + 1) \qquad 0 \le q \le 2, \qquad (11)$$

where $\alpha_D$ is $7/4\pi h^2$ and $21/16\pi h^3$ in 2-D and 3-D respectively.

## 2.2. Governing equations

The governing equations are described by the Navier-Stokes equations for a compressible fluid. The continuity and momentum equations in Lagrangian form can be written as,

$$\frac{d\rho}{dt} = -\rho \nabla \cdot \mathbf{v}, \qquad (12)$$

and

$$\frac{d\mathbf{v}}{dt} = -\frac{1}{\rho} \nabla P + \Gamma + \mathbf{f}, \qquad (13)$$

respectively, where $d$ denotes the total or material derivative, $\mathbf{v}$ is the velocity vector, $\rho$ is density, $P$ is pressure, $\Gamma$ denotes the dissipation terms and $\mathbf{f}$ represents accelerations due to external forces, such as the gravity.

## 2.3. SPH discretisation of the governing equations

The mass conservation property of SPH,

$$\frac{dm}{dt} = 0, \tag{14}$$

where the mass is conserved exactly within a Lagrangian particle, results in a density change due to the volumetric change of the term at the right-hand side of Eq. (12). In the SPH formalism, the discrete form of the continuity equation at a point $a$ with position $\mathbf{r}_a$ reads [14],

$$\left.\frac{d\rho}{dt}\right|_{a \in P} = \rho_a \sum_{b \in P} \frac{m_b}{\rho_b} \mathbf{v}_{ab} \cdot \nabla_a W_{ab} + D_a, \tag{15}$$

where the second term in the right-hand side represents a numerical density diffusion term [21].

The discrete form of the Euler equation in a SPH formalism reads

$$\left.\frac{d\mathbf{v}}{dt}\right|_{a \in F} = -\sum_{b \in P} m_b \left(\frac{P_a + P_b}{\rho_a \rho_b}\right) \nabla W_{ab} + \mathbf{f}, \tag{16}$$

where a symmetric SPH operator has been used that guarantees conservation of momentum [22].

### 2.3.1. Density diffusion terms

Two density diffusion terms (DDT) formulations are implemented in DualSPHysics which act as a high frequency numerical noise filter improving the stability of the scheme by smoothing the density and consequently the pressure. The latter is known to be spurious due to the collocated (in velocity and density) and explicit (time integration) nature of weakly compressible SPH. These terms take the general form of

$$D_a = \delta h c_a \sum_{b \in F} \psi_{ab} \cdot \nabla W_{ab} V_b, \tag{17}$$

where $\delta$ controls the magnitude of the diffusion term. The term $\Psi_{ab}$ is based on the Neumann–Richtmeyer artificial dissipation. Molteni and Colagrossi [23] introduced this artificial dissipation term as

$$\psi_{ab} = 2(\rho_b - \rho_a) \frac{\mathbf{r}_{ab}}{\|r_{ab}\|^2}. \tag{18}$$

Lately, Fourtakas et al. [24] modified Eq. (18) by accounting for the dynamic component of the density by,

$$\psi_{ab} = 2\left(\rho_{ba}^T - \rho_{ab}^H\right) \frac{\mathbf{r}_{ab}}{\|r_{ab}\|^2}, \tag{19}$$

where superscripts $(\cdot)^T$ and $(\cdot)^H$ denote the total and hydrostatic component of the density of a weakly compressible and barotropic fluid by locally constructing the hydrostatic pressure as,

$$P_{ab}^H = \rho_0 g z_{ab}. \tag{20}$$

Although both terms are not consistent near a truncated kernel support [21] (such as free-surface or a wall boundary) thus, resulting in an outwards vector net contribution, the latter term improves the behaviour of pressure near the wall boundaries by operating on the dynamic pressure.

## 2.3.2. Dissipation terms

### 2.3.2.1. Artificial viscosity

An artificial diffusive term can be used in the momentum equation [25] based on the Neumann–Richtmeyer artificial viscosity, aiming to reduce oscillations and stabilise the SPH scheme. The artificial viscosity term is added in the SPH gradient operator in right hand side of Eq. (16) and reads,

$$\Pi_{ab} = \begin{cases} \left(\dfrac{-\alpha \overline{c_{ab}}}{\overline{\rho_{ab}}}\right)\left(\dfrac{h\mathbf{v}_{ab}\cdot\mathbf{r}_{ab}}{r_{ab}^2+\eta^2}\right) & \mathbf{v}_{ab}\cdot\mathbf{r}_{ab} < 0 \\ 0 & \mathbf{v}_{ab}\cdot\mathbf{r}_{ab} \geq 0 \end{cases}, \quad (21)$$

where $(\cdot)_{ab} = (\cdot)_a - (\cdot)_b$ and $\overline{(\cdot)_{ab}} = [(\cdot)_a - (\cdot)_b]/2$. Herein, $c$ denotes the numerical speed of sound (see Section 2.4) and $\alpha$ is the artificial viscosity coefficient. The parameter $\eta = 0.001h^2$, with $\{\eta \in \mathbb{R}; r_{ab} > \eta\}$ guarantees a non-singular operator. It can be shown that $\Pi_{ab} \propto \upsilon_0 \nabla^2 \mathbf{v}$, where $\upsilon_0$ is the kinematic viscosity [14, 26]. The final form of the momentum equation in the presence of artificial viscosity reads,

$$\left.\frac{d\mathbf{v}}{dt}\right|_{a\in F} = -\sum_{b\in P} m_b \left(\frac{P_a + P_b}{\rho_a \rho_b} + \Pi_{ab}\right) \nabla W_{ab} + \mathbf{f}, \quad (22)$$

Due to its simplicity, the artificial viscosity formulation is commonly used in SPH as a viscous dissipation term.

### 2.3.2.2. Laminar viscosity

Viscous dissipation of momentum in the laminar regime in DualSPHysics is approximated by Lo and Shao [27],

$$\upsilon_0 \nabla^2 \mathbf{v}_a = \sum_{b\in P} m_b \frac{4\upsilon_0 \mathbf{r}_{ab} \cdot \nabla W_{ab}}{(\rho_a + \rho_b)(r_{ab}^2 + \eta^2)} \mathbf{v}_{ab}, \quad (23)$$

where $\upsilon_0$ denotes the kinematic viscosity of the fluid.

### 2.3.2.3. Sub-particle scale model

The large eddy simulation sub-particle scale model (SPS) [28] as described by Dalrymple and Rogers [29] using Favre averaging in a weakly compressible approach is implemented in DualSPHysics. The SPS stress tensor $\tau$ is defined in Einstein notation over superscripts $i, j$ according to

$$\tau^{ij} = \overline{v^i v^j} - \overline{v^i}\,\overline{v^j}, \quad (24)$$

modelled by an eddy viscosity closure as,

$$\frac{\tau^{ij}}{\rho} = 2\upsilon_{SPS}\left(S^{ij} - \frac{1}{3}S^{ii}\delta^{ij}\right) - \frac{2}{3}C_l \Delta^2 \delta_{ij} |S^{ij}|^2. \quad (25)$$

Here, $\upsilon_{SPS} = [C_S \Delta]^2 |S^{ij}|^2$, where $C_S = 0.12$ is the Smagorinsky constant $C_l = 0.0066$, $\Delta$ is the initial particle spacing, and $|S^{ij}| = 1/2 \left(2S^{ij}S^{ij}\right)^{1/2}$, where $S^{ij}$ is an element of the SPS strain tensor. Using the same variationally-consistent form of the pressure gradient (Eq. 16), the discrete form of the term reads [29],

$$\frac{1}{\rho}\nabla \cdot \tau_a^{ij} = \sum_{b\in P} m_b \left(\frac{\tau_a^{ij} + \tau_b^{ij}}{\rho_a \rho_b}\right) \nabla^i W_{ab}. \quad (26)$$

Finally, the momentum dissipation term in DualSPHysics takes the following form

$$\Gamma_a = \sum_{b \in P} m_b \frac{4\upsilon_0 \mathbf{r}_{ab} \cdot \nabla W_{ab}}{(\rho_a + \rho_b)(r_{ab}^2 + \eta^2)} \mathbf{v}_{ab} + \sum_{b \in P} m_b \left( \frac{\tau_a^{ij} + \tau_b^{ij}}{\rho_a \rho_b} \right) \nabla^i W_{ab} . \tag{27}$$

## 2.4. Equation of state and compressibility

In the SPH formulation implemented in DualSPHysics, density and pressure are coupled by means of an equation of state (EOS) allowing for weak compressibility of the fluid based on the numerical speed of sound [30],

$$P = \frac{c_s^2 \rho_0}{\gamma} \left[ \left( \frac{\rho}{\rho_0} \right)^\gamma - 1 \right], \tag{28}$$

where $\gamma$ is the polytropic index (usually 7 for water), $\rho_0$ is the reference density, and the numerical speed of sound is defined as $c_s = \sqrt{\partial P/\partial \rho}$ [31].

Therefore, a numerical speed of sound is chosen based on the typical lengthscale and timescale of the domain allowing for a 1% variation in density that results in a Mach number $Ma \approx 0.1$ with, $c_s = 10\|\mathbf{v}\|_{max}$. Such a restriction in the speed of sound allows for larger timesteps within the explicit time integration [30].

## 2.5. Time integrators and time step

In DualSPHysics, two explicit time integration schemes are implemented. For brevity, the governing equations are written as

$$\frac{d\mathbf{v}_a}{dt} = F_a; \quad \frac{d\rho_a}{dt} = R_a; \quad \frac{d\mathbf{r}_a}{dt} = \mathbf{v}_a . \tag{29}$$

The time integrators are briefly introduced below.

### 2.5.1. Verlet time integration scheme

The Verlet scheme [32] is commonly used in molecular dynamics since it is a low computational cost scheme with a second order accurate space integrator that does not require multiple calculation steps within an iteration interval, the WCSPH variables are calculated according to,

$$\begin{aligned} \mathbf{v}_a^{n+1} &= \mathbf{v}_a^{n-1} + 2\Delta t \mathbf{F}_a^n; \quad \mathbf{r}_a^{n+1} = \mathbf{r}_a^n + \Delta t \mathbf{v}_a^n + \tfrac{1}{2}\Delta t^2 \mathbf{F}_a^n; \\ \rho_a^{n+1} &= \rho_a^{n-1} + 2\Delta t R_a^n \end{aligned} \tag{30}$$

Due to the integration over a staggered time interval, the equations of density and velocity are decoupled, which may lead to divergence of the integrated values. Therefore, an intermediate step is required every $N_S$ steps ($N_S \approx 40$ is recommended) according to

$$\begin{aligned} \mathbf{v}_a^{n+1} &= \mathbf{v}_a^n + \Delta t \mathbf{F}_a^n, \quad \mathbf{r}_a^{n+1} = \mathbf{r}_a^n + \Delta t \mathbf{v}_a^n + \tfrac{1}{2}\Delta t^2 \mathbf{F}_a^n, \\ \rho_a^{n+1} &= \rho_a^n + \Delta t R_a^n \end{aligned} \tag{31}$$

where the superscript $n \in \mathbb{N}$ and denotes the time step and $t = n\Delta t$.

### 2.5.2. Symplectic position Verlet time integration scheme

The symplectic position Verlet time integrator scheme [33] is second order accurate in time. It is ideal for Lagrangian schemes as it is time reversible and symmetric in the absence of diffusive terms that preserve geometric futures. The position Verlet scheme in the absence of dissipation forces reads,

$$\begin{aligned}
\mathbf{r}_a^{n+1/2} &= \mathbf{r}_a^n + \frac{\Delta t}{2}\mathbf{v}_a^n; \\
\mathbf{v}_a^{n+1} &= \mathbf{v}_a^n + \Delta t \mathbf{F}_a^{n+1/2}; \\
\mathbf{r}_a^{n+1} &= \mathbf{r}_a^{n+1/2} + \frac{\Delta t}{2}\mathbf{v}_a^{n+1}.
\end{aligned} \tag{32}$$

However, in the presence of viscous forces and density evolution in DualSPHysics, the velocity is required at the $(n + 1/2)$ step thus, a velocity Verlet half step is used to compute the required velocity for the acceleration and density evolution for $\mathbf{F}(\mathbf{r}_{n+1/2})$ and $R(\mathbf{r}_{n+1/2})$, respectively. The scheme implemented in DualSPHysics reads,

$$\begin{aligned}
\mathbf{r}_a^{n+1/2} &= \mathbf{r}_a^n + \frac{\Delta t}{2}\mathbf{v}_a^n \\
\mathbf{v}_a^{n+1/2} &= \mathbf{v}_a^n + \frac{\Delta t}{2}\mathbf{F}_a^n \\
\mathbf{v}_a^{n+1} &= \mathbf{v}_a^n + \Delta t \mathbf{F}_a^{n+1/2} \\
\mathbf{r}_a^{n+1} &= \mathbf{r}_a^n + \Delta t \frac{\left(\mathbf{v}_a^{n+1} + \mathbf{v}_a^n\right)}{2}
\end{aligned}, \tag{33}$$

where $\mathbf{r}^{n+1/2}$ is substituted to $\mathbf{r}^{n+1}$ in Eq. (32) to eliminate dependence from $\mathbf{u}^{n+1/2}$. Finally, the density evolution follows the half time steps of the symplectic position Verlet scheme as follows [34],

$$\begin{aligned}
\rho_a^{n+1/2} &= \rho_a^n + \frac{\Delta t}{2}R_a^n \\
\rho_a^{n+1} &= \rho_a^n \frac{2 - \varepsilon_a^{n+1/2}}{2 + \varepsilon_a^{n+1/2}}
\end{aligned}, \tag{34}$$

where $\varepsilon_a^{n+1/2} = -\left(R_a^{n+\frac{1}{2}}/\rho_a^{n+1/2}\right)\Delta t$.

### 2.5.3. Variable time step

The time integration is bounded by the Courant–Friedrich–Levy (CFL) condition necessary in explicit time integration schemes to limit the numerical domain to the physical domain of dependence [35],

$$\Delta t_f = \min_a\left(\sqrt{h/|f_a|}\right); \quad \Delta t_{cv} = \min_a \frac{h}{c_s + \max_b \frac{|h\mathbf{v}_a \cdot \mathbf{r}_a|}{r_{ab}^2 + \eta^2}}, \tag{35}$$

$$\Delta t = C_{CFL}\min\left(\Delta t_f, \Delta t_{cv}\right)$$

where $|f_a|$ is the force per unit mass from Eq. (16). The variable time step is chosen as the minimum between $\Delta t_f$ and $\Delta t_{cv}$, and is bounded by the Courant number $C_{CFL}$, usually in the range of 0.1 to 0.2.

## 2.6. Boundary conditions

Several boundary conditions are implemented in DualSPHysics. Solid boundary conditions are discretised by a set of boundary particles $a \in B$ that differ from the fluid particles F. Due to the Lagrangian nature of SPH, solid boundaries can also be moved straightforwardly according to user-defined motion. The two solid boundary conditions currently implemented in DualSPHysics are briefly described below.

### 2.6.1. Dynamic boundary conditions

The dynamic boundary condition (DBC) discretises a solid wall with a set of boundary particles B for which the continuity equation is evaluated as [36]

$$\left. \frac{d\rho}{dt} \right|_{a \in B} = \rho_a \sum_{b \in F} \frac{m_b}{\rho_b} \mathbf{v}_{ab} \cdot \nabla W , \qquad (36)$$

while the position of these particles is updated following the equation

$$\left. \frac{d\mathbf{v}}{dt} \right|_{\forall a \in B} = F_{\text{(imposed)}} , \qquad (37)$$

where $F_{\text{(imposed)}}$ is the force imposed on moving boundary particles by the fluid, an external force or another solid object as described in Sections 3.2 and 4. A motion can be imposed by predefined user function or accelerations integrated in time. The increase in density that arises from Eq. (36) results in a pressure increase in the momentum equation of particle $a \in F$, yielding a repulsive force between the fluid and boundary particles. This simplified boundary condition is ideal for GPU implementation since it enables code optimisation using vector lists. More details on this formulation can be found in Crespo et al. [36].

### 2.6.2. Modified dynamic boundary conditions

Starting from version 5.0 of DualSPHysics, an alternative way to discretize walls and other impermeable boundaries has been introduced to overcome some of the limitations of the DBC approach. This formulation has been named Modified-DBC (mDBC). Boundary particles used in mDBC are arranged in the same way as DBC particles. However, for each mDBC boundary particle, a ghost node is projected into the fluid across a boundary interface, similarly as the procedure employed by Marrone et al. [37]. Fluid properties are then computed at the ghost node through a corrected SPH approximation proposed by Liu and Liu [38] and finally mirrored back to the boundary particles. More details on this boundary condition can be found in English et al. [39].

The fluid density and its gradient are computed at the ghost node, and then the density of the boundary particle $\rho_B$ is obtained by means of a linear approximation in the form:

$$\rho_B = \rho_G + (\mathbf{r}_B - \mathbf{r}_G) \cdot \langle \nabla \rho_G \rangle , \qquad (38)$$

where $r_B$ and $r_G$ are the position of the boundary particle and associated ghost node, respectively, with $B \cap G$, and $\langle \nabla \rho_g \rangle$ is the corrected SPH gradient at the ghost node. The velocity at the ghost node is found using a Shepard filter sum

$$\mathbf{v}\big|_{a\in G} = \frac{\sum_{b\in F} V_b \mathbf{v}_b W}{\sum_{b\in F} V_b W}, \qquad (39)$$

The boundary particle then receives this velocity with the direction reversed to create a no-slip condition at the boundary interface. With this method, it is also possible to create a free-slip boundary by assigning the exact tangential velocity found at the ghost node to the boundary particles.

### 2.7. Periodic boundary conditions

Periodic boundary conditions (PBC), which can be used to describe an infinitely long domain by a finite cell domain, are implemented in DualSPHysics. This is achieved by allowing particles located within $kh$ (with $k = 2$ for the smoothing kernels implemented in DualSPHysics) near an open lateral boundary to interact with the fluid particles of the opposite lateral boundary completing the truncated support in a cyclic manner in 2-D and 3-D.

### 2.8. Particle shifting algorithm

The anisotropic distribution of particles introduces additional discretisation error through the zeroth and higher order kernel moments (i.e. the discrete version of Eq. (3)). This is especially true in negligible or large dynamics [30] and violent flows where particles may not maintain an isotropic distribution. In DualSPHysics, the Fickian based particle shifting algorithm of Lind et al. [40] is used to maintain a near isotropic particle distribution,

$$\delta \mathbf{r}_a\big|_{a\in F} = -D_s \nabla C_a, \qquad (40)$$

where $\delta \mathbf{r}_a$ is the shifting distance, $\nabla C_a$ is the kernel gradient and $D_s$

$$D_s = A_s h \|\mathbf{v}_a\| \Delta t, \qquad (41)$$

where $A_s$ is a parameter and $\|\mathbf{v}_a\|$ is the velocity magnitude of particle $a$ [41]. This imposes a restriction based on the particle velocity magnitude to prevent excessive movement and loss of information from particles moving to different domain cells. The parameter value $A_s$ is in the range $[1-6]$, with a value of 2 recommended by Skillen et al. [41]. Expressions for $\nabla C_a$ are given in Section 3.4

## 3. Functionalities

### 3.1. Wave generation and absorption in DualSPHysics

DualSPHysics employs moving boundary particles to generate regular and random long-crested waves. Moving boundaries mimic wavemakers typical of experimental facilities, namely piston-type and flap with variable draft. Moreover, a piston wavemaker allows solitary wave generation. Active and passive wave absorption techniques are implemented to allow the generation of long time wave series in relatively short domains with negligible reflection. Wave absorption techniques can be categorized as passive and active, respectively. Passive wave absorbers are usually required to damp the wave energy and to reduce the reflection exerted by the boundary of the model domain. Further details on passive absorption in DualSPHysics can be found in Altomare et al. [42]. In active absorption systems, the wavemaker real-time displacement is corrected to cancel out the reflected waves and to damp the re-reflection phenomenon altogether. For a complete description of the implementation, the reader can refer to Altomare et al. [42] and Domínguez et al. [43].

*3.1.1. Generation of regular waves*

The Biesel transfer functions [44] relate the wavemaker displacement to the water surface elevation. Based on Hughes [45], the general solution for the second-order wavemaker motion to suppress second harmonics can be expressed as follows:

$$X_0(t) = \frac{H}{2m_{cr}}\sin\left(\frac{2\pi t}{T}+\phi\right) + \left[\left(\frac{H^2}{32\left(1-\frac{d}{2(d+d_0)}\right)}\right)\cdot\left(\frac{3\cosh\left(\frac{2\pi d}{L}\right)}{\sinh^3\left(\frac{2\pi d}{L}\right)} - \frac{2}{m_{cr}}\right)\right]\sin\left(\frac{4\pi t}{T}+2\phi\right), \quad (42)$$

where $m_{cr}$ is given by

$$m_{cr} = \frac{4\sinh\left(\frac{2\pi d}{L}\right)}{\sinh\left(\frac{4\pi d}{L}\right)+\frac{2\pi d}{L}}\left[\sinh\left(\frac{2\pi d}{L}\right)+\frac{1-\cosh\left(\frac{2\pi d}{L}\right)}{\frac{2\pi}{L}(d+d_0)}\right], \quad (43)$$

where $H$ is the wave height, $d$ is the initial water depth, $\varphi$ [0:2$\pi$] is the initial phase, $x$ is distance along the wave propagation and $d_0$ is the location of the wavemaker hinge (for flap: $d_0 < 0$ if hinge is above the sea bottom, $d_0 > 0$ if below; for piston $d_0 \rightarrow \infty$). The quantity $2\pi/T$ is the angular frequency and $2\pi/L$ corresponds to the wave number. $T$ is the wave period and $L$ the wave length. Eqs. (42) and (43) are valid for regular waves and for flap- and piston-type wavemakers. $X_0(t)$ is the wavemaker displacement at the free surface: for a piston, this corresponds to the horizontal movement of the entire board, $e(t)$; for a flap, the amplitude of the motion is expressed by the angle swept in time around the hinge, and calculated as $\Theta(t) = \tan^{-1}(X_0(t)/(d+d_0))$. The application of Eq. (42) is limited to values of the Ursell number $HL^2/d^3 < 8\pi^2/3$.

*3.1.2. Generation of random waves*

Irregular wave generation is performed in DualSPHysics based on Liu and Frigaard [46]. Starting from an assigned wave spectrum, the transfer function is applied to each component in which the spectrum is discretised. Composing all the *i*-th components of the wavemaker displacement will provide the first-order solution for irregular waves. Two standard wave spectra are implemented in DualSPHysics: JONSWAP and Pierson-Moskowitz spectra. The characteristic parameters of each spectrum can be assigned by the user together with the value of $N_s$ (number of parts into which the spectrum is divided). The spectral frequency discretization can be done in four different ways: equidistant discretization, unevenly distributed, and by applying a stretched or cosine stretched function. The use of the last two methods has shown the most accurate results in terms of wave height distribution and groupness, even for low values of $N$.

Second-order wave generation is implemented in order to cancel out the parasitic long waves [47] and the spurious displacement long waves, the latter ones caused by finite wavemaker displacements away from the mean position. The method implemented in DualSPHysics is based on the solution for the control signal of the wavemaker that is described in Barthel et al. [48] and is fully described in Altomare et al. [42].

### 3.1.3. Solitary waves

The displacement of a piston-type wavemaker to generate a solitary wave is prescribed by the following equation [49]:

$$x_s(t) = \frac{2H}{Kd} \tanh\left[K(ct - x_s(t))\right], \tag{44}$$

in which the parameter $K$ is the outskirt coefficient describing the way the free surface elevation tends towards the mean level at infinity. Three different theories for solitary wave generation have been implemented in DualSPHysics. The three different approaches result in different wave propagation properties, i.e. different loss of amplitude and differences in generated trailing waves [43]. One solution is derived from the Rayleigh solitary wave solution; the second one uses a wavemaker law of motion derived from the KdV solution; the third one consists of solving Eq. (44) iteratively using the Newton-Raphson method.

Additionally, DualSPHysics allows the generation of a series of multiple solitary waves where the motion of the wavemaker is calculated for each $i$-th solitary wave. Any of the three aforementioned generation theories can be used. Then, the time lag between the ($i$+1)-th and $i$-th solitary waves has to be specified as a fraction or multiple of the generation time used for the $i$-th solitary wave. The generation of multiple solitary waves is a simple but efficient way to model a wave train of several tsunamis that might have been triggered by the same tectonic event. For further details on solitary wave generation in DualSPHysics, please see Domínguez et al. [43].

### 3.1.4. Active wave absorption

The implemented active wave absorption system is based on the time-domain filtering technique that uses the free-surface elevation at the wavemaker position as feedback for the control of the wavemaker displacement. The wavemaker velocity is corrected to match the velocity induced by the wave that will be absorbed. For a piston-type wavemaker, the wave absorption is performed using the linear long wave theory [50]. The corrected velocity, $U_c$, is expressed as follows:

$$U_C(t) = U_i(t) - (\eta_{WG,SPH}(t) - \eta_{AWAS,i}(t))\sqrt{g/d}, \tag{45}$$

where $\eta_{AWAS,i}$ is the target incident free-surface elevation and $\eta_{WG,PH}$ is the measured one in front of the wavemaker, $U_i$ is the theoretical wave maker velocity, equal to the derivative in time of the piston-type wavemaker displacement $e(t)$. The wavemaker position at $t + \Delta t$ is then corrected at the end of the time step using the following equation:

$$e(t + \Delta t) = e(t) + \left(U_C(t + \Delta t) + U_C(t)\right)\frac{\Delta t}{2}. \tag{46}$$

## 3.2. Fluid-driven objects

One of the most interesting capabilities of SPH models is the simulation of fluid-driven objects. The implementation of this functionality in DualSPHysics derives the motion of a floating structure by considering its interaction with fluid particles and using the resulting forces to drive its motion.

First, the net force on each individual particle that form the same floating object is computed as a summation of the contributions of all surrounding fluid particles. In this way, each floating particle $a \in K$ experiences an acceleration ($d\mathbf{v}/dt$) given by

$$\left.\frac{d\mathbf{v}}{dt}\right|_{a\in K} = \sum_{b\in F} m_b \frac{d\mathbf{v}_{ab}}{dt}. \tag{47}$$

As the object is assumed rigid, the basic equations of rigid body dynamics are solved to obtain its motion:

$$M\frac{d\mathbf{v}}{dt} = \sum_{b\in K} m_b \frac{d\mathbf{v}_b}{dt}, \tag{48}$$

$$I\frac{d\mathbf{\Omega}}{dt} = \sum_{b\in K} m_b \left(\mathbf{r}_b - \mathbf{R}_0\right)\frac{d\mathbf{v}_b}{dt}, \tag{49}$$

where $M$ is the total mass of the object, $I$ the moment of inertia, $\mathbf{v}$ the velocity, $\mathbf{\Omega}$ the angular velocity and $\mathbf{R}_0$ the centre of mass. Eqs (48) and (49) are integrated in time to obtain the values of $\mathbf{v}$ and $\mathbf{\Omega}$ at the beginning of the subsequent time step. Each particle belonging to the object will move with a velocity $\mathbf{v}_k$ given by

$$\mathbf{v}_k = \mathbf{v} + \mathbf{\Omega} \times \left(\mathbf{r}_k - \mathbf{R}_0\right). \tag{50}$$

This technique has been shown to preserve both linear and angular momenta [51].

### 3.3.  Open boundaries (inlet & outlet)

Several types of open boundary approaches are available in SPH, among which the unified semi-analytical model [52–54], the mirror particles method [55], and the buffer region strategy, originally introduced by Lastiwka et al. [56], then used in many other works [57–60]. The latter approach has been chosen for implementation in DualSPHysics, mainly for its simplicity but also for its suitability to a variety of engineering applications.

The rationale behind using buffer regions is to attach several layers of fictitious SPH particles to a region of the computational domain where an open boundary is needed, for example an inlet or an outlet. Buffer particles are usually deployed to follow the geometry of the open boundary. Figure 3 depicts a buffer region near a permeable boundary, where an inflow (or outflow) of SPH particles is desired. Buffer particles are extruded starting from the permeable boundary, i.e. the fluid-buffer interface, with a finite number of layers chosen to equal or exceed the kernel radius, ensuring full kernel support for the closest fluid particles to the inlet/outlet threshold. Once the buffer particles are created, it is necessary to enforce boundary conditions appropriately. This can be done by assigning physical quantities to the buffer particles in two different ways: the first is to simply pre-assign the quantities to these particles, for example a certain velocity or pressure field defined by the SPH user. The second option is to extrapolate the physical quantities from the fluid domain, using ghost nodes properly placed within the fluid particles neighbouring the open boundary. The placement of ghost nodes is shown in Figure 3 for three different buffer particles, although it is noted that every buffer particle must be assigned a corresponding ghost node in the fluid. Ghost nodes are created by mirroring the position of the buffer particles into the fluid domain, following the normal direction to the permeable boundary.

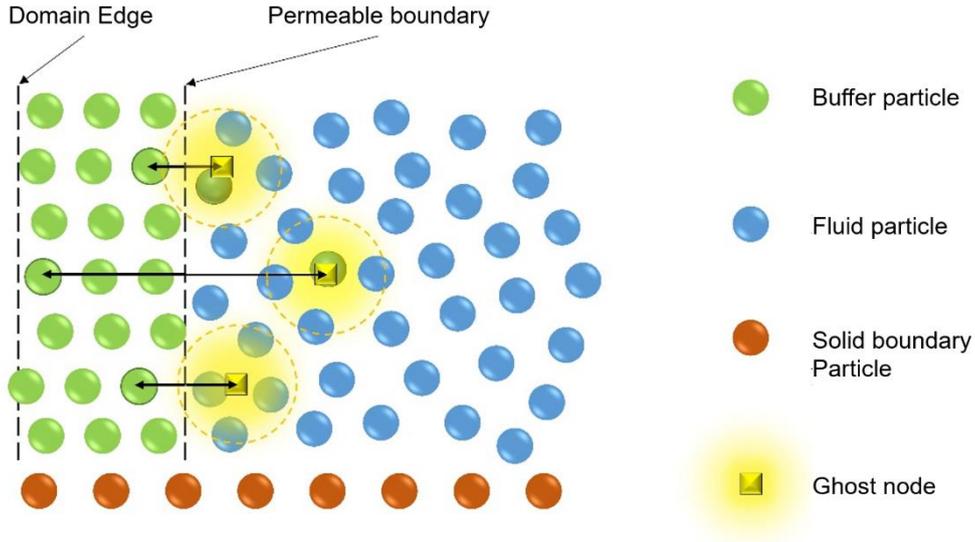

Figure 3. A sketch of the placement of ghost nodes for the open boundary algorithm implemented in DualSPHysics.

When extrapolating fluid properties from the ghost nodes, the idea for retrieving quantities of the buffer particles is to perform an SPH interpolation at the ghost nodes, and then transport that information back to the buffer particles. However, a standard SPH particle interpolation would not be consistent because these nodes are situated near an open boundary, and thus not fully surrounded by fluid particles. To obviate this loss of consistency, the algorithm proposed in Liu and Liu [38] is implemented, restoring first order kernel and particle consistency. More details about the fluid extrapolation via ghost nodes in DualSPHysics, including the linear system solved at each time step, can be found in Tafuni et al. [61].

The implementation of open boundaries in DualSPHysics is done in a versatile way, with no distinction between an inlet and an outlet region: a buffer area can behave simultaneously as an inlet and outlet, enabling the use of DualSPHysics to problems with backflow and flow reversion, see for example [62]. Other attractive features for use in real engineering problems include: (a) the capability of enforcing unsteady velocity and pressure profiles and/or pressure and velocity gradients along a given direction; (b) the simulation of variable free-surface elevation, an essential aspect in flow with a free surface because of the presence of waves entering and exiting the computational domain; (c) the ability to generate several buffer geometries, including square, circular and other common engineering shapes.

### 3.4. Multi-phase: Liquid and gas

In its original release, the DualSPHysics code could only deal with single-phase flows. This limits the range of applications, as a diverse range of problems such as liquid-gas flows with mixing and violent free-surface hydrodynamic interaction cannot be accurately modelled. To address these and other problems typical of industrial and research fields, such as coastal and nuclear engineering, DualSPHysics includes the multi-phase model from Mokos et al. [63].

The multi-phase model uses a modified version of Tait's equation of state [31] (the same equation used in single-phase DualSPHysics) for both liquids and gases, with an additional term representing the cohesion forces between particles of the same phase as proposed by Nugent and Posch [64].

$$P = \frac{c_s^2 \rho_0}{\gamma}\left[\left(\frac{\rho}{\rho_0}\right)^\gamma - 1\right] + X - \alpha\rho^2, \text{with } \alpha = 1.5g\left(\frac{\rho_l}{\rho_g^2}\right)L_c, \tag{51}$$

where *X* represents a constant background pressure. The last term is only used for the gas phase and the cohesion coefficient $\alpha$ depends on the properties of the different phases ($\rho_l$ and $\rho_g$ are the initial densities of the liquid and the gas) and a characteristic length scale $L_c$ of the problem [65].

For the continuity and momentum equations, the multi-phase model uses the variationally consistent forms of the velocity divergence and pressure gradient as proposed by Colagrossi and Landrini [66]. An additional term, originating from the cohesion term in the equation of state, is included in the momentum equation for gases [64].

$$\left.\frac{d\mathbf{v}}{dt}\right|_{a \in F} = -\sum_{b \in P} m_b \left( \frac{P_a + P_b}{\rho_a \rho_b} + \Pi_{ab} \right) \nabla W_{ab} + 2\alpha \rho_a^2 \sum_{b \in P} \frac{m_b}{\rho_a \rho_b} \nabla W_{ab} + \mathbf{g} . \tag{52}$$

The use of high resolution, made possible by GPUs, allowed the identification of a numerical inconsistency for phases with a high density ratio, particularly in violent flows. Unphysical voids and phase separation occur, ultimately leading to numerical instability. The Fickian-based particle shifting algorithm of Section 2.8, with a selectively activated free-surface correction, has been adapted for multi-phase to prevent the creation of unnatural voids and maintain numerical stability through nearly uniform distributions [63]. The Fickian shifting algorithm is a modified version of the one developed by Lind et al. [40], adapted for multi-phase flows. The algorithm uses Fickian diffusion to shift particles towards areas with lower concentration, while limiting the movement on the normal direction to the free surface. The particle concentration gradient is computed through an SPH gradient approximation with the particle concentration *C* computed through the sum of the smoothing kernel.

$$\nabla C|_{a \in F} = \sum_{b \in P} \frac{m_b}{\rho_b} (C_b - C_a) \nabla W_{ab} \text{ with } C_a = \sum_{b \in P} \frac{m_b}{\rho_b} W_{ab} . \tag{53}$$

The concentration gradient allows the computation of the particle shifting distance, $\delta r_s$, with the Fickian diffusion coefficient computed through the method of Skillen et al. [41].

To prevent the unphysical movement at the free surface Lind et al. [40] proposed that the concentration gradient near the surface was controlled using the local tangent (**s**) and normal (**n**) vectors at the free surface. The free surface equation is expanded to 3-D so that the bitangent vector (**b**, orthogonal to **s** and **n**) is also considered [63].

$$\delta \mathbf{r}_s = -D_s \left( \frac{\partial C_a}{\partial s} \mathbf{s} + \frac{\partial C_a}{\partial b} \mathbf{b} + a_n \left( \frac{\partial C_a}{\partial n} - \beta_n \right) \mathbf{n} \right), \tag{54}$$

where $\beta_n$ is the initial concentration gradient and $\alpha_n$ is a numerical parameter limiting diffusion in the normal direction. This is set to 0 for violent flows, and 0.1 for long slow flows, such as standing gravity waves [40]. For the multi-phase flow, the free surface term is applied to the liquid-gas interface, but only on the liquid particles. Shifting for the gas phase should not be restricted in the interface, to allow for free expansion to areas of low concentration, similar to the behaviour of an actual gas [63].

For flows in which the Weber or the Eötvös number is sufficiently small, surface tension forces need to be taken into account. In order to model these flows, DualSPHysics includes a surface tension formulation. The method uses the Continuum Surface Force (CSF) model [67] and the SPH implementation by Hu and Adams [68]. This implementation uses a tensor formulation to avoid computing the surface curvature as proposed by Lafaurie et al. [69]. A colour function is used to distinguish between the different phases and compute the surface normal [70].

### 3.5. Multi-phase: Non-Newtonian flows

The rheology of materials such as polymers, slurries, pastes and suspensions is a frequent occurrence in industrial applications and academic research. These materials are usually characterized by a yield stress below which the material has not yielded and does not flow. Furthermore, the shear strain to shear stress relationship may not be linear as with Newtonian fluids. In general, fluids that do not obey the Newton law of viscosity may be classified as non-Newtonian fluids with viscoelastic, time dependant viscosity and non-Newtonian viscosity.

In DualSPHysics the generalized Herschel-Bulkley-Papanastasiou [71, 72] (HBP) model has been implemented following the work of Fourtakas and Rogers [73] on liquid-sediment two-phase flows. The HBP approach allows the modelling of a wide range of viscoplastic materials, such as Bingham plastic, Bingham pseudoplastic and dilatant fluids. Furthermore, the HBP generalised non-Newtonian model can be used to simulate shear thinning or thickening materials in the absence of yield strength. The constitutive equation for the HBP model is written as,

$$\tau^{ij} = \eta_{app}\dot{\gamma}^{ij}, \tag{55}$$

where $\tau^{ij}$ is the shear stress tensor and $ij$ denote coordinate directions using the Einstein notation. By dropping the Einstein notation for brevity, the symmetric shear rate tensor $\dot{\gamma}$ given by

$$\dot{\gamma} = \left[\nabla \mathbf{v} + \nabla \mathbf{v}^T\right], \tag{56}$$

with $\nabla \mathbf{v}$ and $\nabla \mathbf{v}^T$ representing the velocity gradient and its transpose, respectively, while $\eta_{app}$ is the apparent viscosity

$$\eta_{app} = \mu|\dot{\gamma}|^{n_{HB}-1} + \frac{\tau_y}{|\dot{\gamma}|}\left[1 - e^{-m_P|\dot{\gamma}|}\right]. \tag{57}$$

In the above, $\tau_y$ is the yield stress of the fluid, $\mu$ is the viscosity (or consistency index), $n_{HB}$ is the power law index (commonly referred to as the Herschel-Bulkley parameter), $m_p$ is the exponential shear rate growth parameter (or Papanastasiou parameter), and $|\dot{\gamma}|$ is the magnitude of the symmetric strain rate tensor defined as

$$|\dot{\gamma}| = \sqrt{\frac{1}{2}\left((tr(\dot{\gamma}))^2 - tr(\dot{\gamma}^2)\right)}. \tag{58}$$

The Papanastasiou parameter controls the growth of the stress with a finite value for small shear rates in the unyielded region, and the linear behaviour in the yielded region. Note that, as $m_p \to \infty$, the HBP model reduces to the original Herschel-Bulkley model, and with $n_{HB} = 1$ it further reduces to a simple Bingham model. When $n_{HB} = 1$ and $m_p = 0$, the model reduces to a Newtonian fluid. Moreover, the HBP model does not exhibit a discontinuity at zero shear rates, in contrast with a pure Bingham model, which is usually discretized using a bi-viscosity model.

The implementation of the HBP model in DualSPHysics is outlined next. The momentum Eq. (13) is written as

$$\left.\frac{d\mathbf{v}}{dt}\right|_{a \in F} = -\sum_{b \in P} m_b\left(\frac{P_b + P_a}{\rho_b\rho_a}\right)\nabla_a W_{ab} + \left\langle\frac{1}{\rho}\nabla \cdot \tau\right\rangle + \mathbf{g}, \tag{59}$$

where the brackets denote an SPH approximation. Two formulations are available in the multi-phase non-Newtonian formulation for the discretization of the viscous forces. The first uses the formulation of Morris et al. [74] that reads

$$\left.\frac{1}{\rho}\nabla \cdot \tau\right|_{a \in F} \approx \upsilon_{app}\nabla^2 \mathbf{v}_a = -\sum_{b \in P}\frac{m_b(\upsilon_a + \upsilon_b)\mathbf{r}_{ab} \cdot \nabla_a W_{ab}}{\rho_b(|r_{ab}|^2 + 0.001h^2)}\mathbf{v}_{ab}, \tag{60}$$

where $v_{app}$ is the apparent kinematic viscosity. The second formulation takes advantage of the conservative gradient operator in SPH form and the viscous forces are discretised according to Eq. (26).

The continuity Eq. (12) has the additional density ratio terms for the interpolating and neighbouring particles, allowing different densities to be used in each phase of the system. It is thus written as

$$\left.\frac{d\rho}{dt}\right|_{a\in F} = \rho_a \sum_{b\in P} \frac{m_b}{\rho_b} \mathbf{v}_{ab} \cdot \nabla_a W_{ab} + D_{a\in M}, \tag{61}$$

with $M = \sqcup_{n\in \mathbb{N}} M_n$, and $n$ is the number of phases. For the discretisation of the shear stress tensor and velocity gradients of Eq. (56), a finite-difference approach or an SPH gradient summation can be used. The former reads

$$\nabla \mathbf{v}_{ab} = \frac{\mathbf{v}_{ab} \cdot \mathbf{r}_{ab}}{|r_{ab}|^2}, \tag{62}$$

whereas the latter takes the following form

$$\nabla \mathbf{v}|_{a\in P} = \sum_{b\in P} \frac{m_b}{\rho_b} (\mathbf{v}_a - \mathbf{v}_b) \nabla_a W_{ab}. \tag{63}$$

In this implementation, any combination of viscous operators and gradients is allowed in the multi-phase non-Newtonian DualSPHysics formulation. More information about the multi-phase non-Newtonian formulation can be found in Fourtakas and Rogers [73]. Future implementations are planned, and include a sediment phase with yielded and unyielded sediment regions, as described in the aforementioned manuscript.

## 4. Coupling with other models

Following the guidelines in the "SPHERIC Grand Challenge #4: Coupling to other models", it has been concluded that, depending on the problem under scope, it can be much more efficient to couple SPH with a different numerical method rather than sticking to a single method of choice. The main reason for coupling is to enhance the capabilities of both models beyond their specific application fields, thus enabling the study of a wider range of problems. This section describes the successful coupling strategies of DualSPHysics with other models. For brevity, the governing equations of the coupled models are not included below, but details and reference literature are provided.

### 4.1. Coupling with wave propagation models

Multi-scale complex simulations, often characterized by three-dimensional domains and long durations, require extensive computational resources if only one stand-alone model is employed. In the special case of coastal engineering applications, the proper simulation of all processes from wave generation to wave-structure/beach interaction is required. SPH-based solvers allow a detailed model of local phenomena, such as interaction between sea waves and coastal defences [75, 76]. The mesh-free nature of SPH is successful at mimicking the wave generation system of experimental facilities, employing moving boundaries such as piston- or flap-type wavemakers. However, when using these schemes, waves need to be generated and propagate over relatively long domains to simulate wave transformation properly and thus guarantee the correct wave sea state at the toe of the structure, similarly to what happen in physical model tests. To reduce the computational cost of such simulations, less computationally demanding wave propagation models can be coupled with SPH-based solvers.

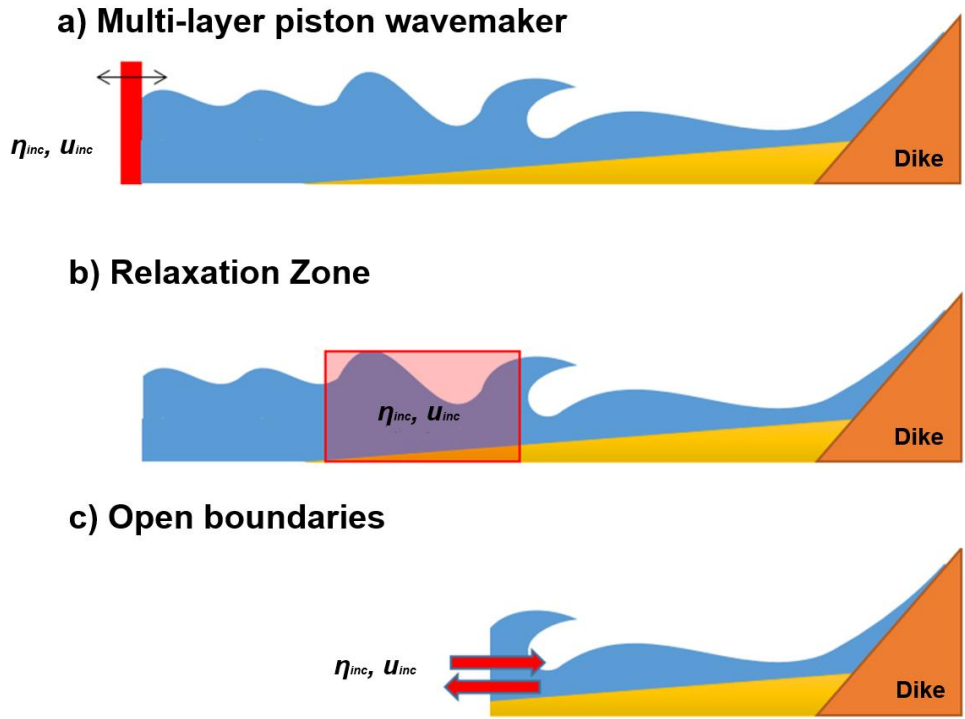

Figure 4. Representation of the coupling techniques between DualSPHysics and wave propagation models: $u_{inc}$ and $\eta_{inc}$ are the velocity and water surface elevation from the wave propagation model used as steering signal in DualSPHysics.

The three coupling schemes between the DualSPHysics and SWASH models shown in Figure 4 are hereby listed and briefly described:

- Multi-layered piston: moving boundaries are employed to generate waves. This scheme has been coupled with the SWASH model [77] by Altomare et al. [78]. The real time displacement of the multi-layered piston wavemaker is reconstructed on the basis of the velocity time series calculated by the SWASH model and interpolated along the water depth at a specific $x$-coordinate (here the $x$-axis corresponds to the direction of wave propagation). There is no reflection compensation, therefore the use of this technique is recommended for low-reflective cases or very short time series. It is a one-way offline coupling and at least 8 layers are recommended to be used in SWASH. The same strategy was applied to develop a two-way coupling with a fully nonlinear potential flow solver OceanWave3D [79] by Verbrugghe et al. [80]. In this case, both models (DualSPHysics and OceanWave3D) exchange sea-state information throughout the simulation.

- Relaxation zone [81]: the coupling zone is extended from one single location to an area. In this one-way offline coupling, the orbital velocity calculated by the wave propagation model is imposed to the fluid particles in DualSPHysics. The particle velocity is controlled by an analytical solution with a weighting function $C$ to balance the reflected waves inside the relaxation zone. A correction for the velocity of fluid particles is implemented to avoid unrealistic increase of the water level outside the generation zone, caused by Stokes drift velocity.

- Open boundaries: an improved two-way coupling with OceanWave3D is presented in Verbrugghe et al. [82]. The open boundary algorithm in Section 3.3 is employed at the coupling interfaces. At the inlet, horizontal orbital velocities and surface elevations are imposed on the buffer particles. At the outlet, horizontal orbital velocities are imposed, while surface elevation is extrapolated from the SPH fluid domain. A velocity correction algorithm based on linear shallow water theory is applied for absorbing wave reflection in the SPH fluid domain, similar to the one described in Altomare et al. [42]. The SPH surface elevation is coupled back to OceanWave3D to replace the original signal.

All coupling methodologies described above used 2-D tests for validation, proving a model accuracy that is generally in the order of the smoothing length. Inherently, open boundaries allow coupling in shallow waters, where wave non-linearity and mass transport are dominant (e.g. in wave breaking zone).

## 4.2. Coupling with Discrete Element Method (DEM)

Simple (i.e. granular-type) simultaneous solid–fluid and solid–solid interactions represent a set of problems common to several engineering disciplines such as coastal, offshore, maritime and fluvial engineering. In order to characterize such flows, DualSPHysics includes a Distributed Contact Discrete Element Method (DCDEM) [83] within its SPH formulations for fluid-solid modelling. The general concept is to compute the forces acting on a fluid-solid particle pair using the SPH formulation and the solid-solid interactions via DEM, retaining the same explicit integrator and also the DualSPHysics meshless framework. The DCDEM concept introduced by Cummins and Cleary [84] is used, allowing arbitrary solid shapes to be reproduced by a collection of particles that are then subjected to a rigid body restriction. Given adequate non-linear contact models [85], a wide range of material behaviours can be modelled using three basic parameters: Poisson ratio, Young's modulus and dynamic friction coefficient. The drawbacks of the model are common to all general DEM formulations: the explicit nature may impose very small stability regions, extended frictional contacts are either computationally expensive or inaccurately reproduced and low resolution simulations may lead to artificial geometry effects due to the underlying spherical shape of the particles.

Accounting for these limitations, given its simplicity and speed, applications include assessment of the severity of hydrodynamic actions on structures, risk of natural hazards, design of floating bodies or design of exposed structures.

## 4.3. Coupling with Project Chrono

Given the shortcomings of the DCDEM approach for general simulations, the Differential Variational Inequality (DVI) implementation of Project Chrono is leveraged [86]. The implementation resorts to linearising and efficiently solving a set of Cone Complementary Problems describing the restricted dynamics of a multi-body system. The notion of restriction is loose, so besides non-penetration contacts, mechanical-type restrictions can be imposed, such as joints, hinges, sliders, springs and combinations of these. Given the underlying mesh representing the bodies (that DualSPHysics generates and maps to an underlying SPH particle distribution), frictional contacts are fully described, including sliding and rolling friction. Thanks to the efficiency of the DVI formulation, the impact on the computational time is residual, allowing the simulation of large and complex multiphysics systems. In addition, the DVI formulation presents several advantages comparing with explicit DEM models since it has a much larger stability region leading to larger time-steps.

The work of Canelas et al. [87] included details of the implementation and validation with fluid–structure–structure interaction cases. Moreover, several hypothetical cases were also presented to demonstrate the capabilities of DualSPHysics coupled with Project Chrono.

Figure 5 shows the flow chart of the DualSPHysics coupling with Project Chrono. It can be seen how DualSPHysics computes the linear and angular acceleration of floating objects starting from interaction with fluid particles. These values and the *dt* are passed to Project Chrono, who calculates final position and velocities according to the constraints defined for Chrono model. Finally, DualSPHysics updates the data of the floating particles according to the information received from Project Chrono.

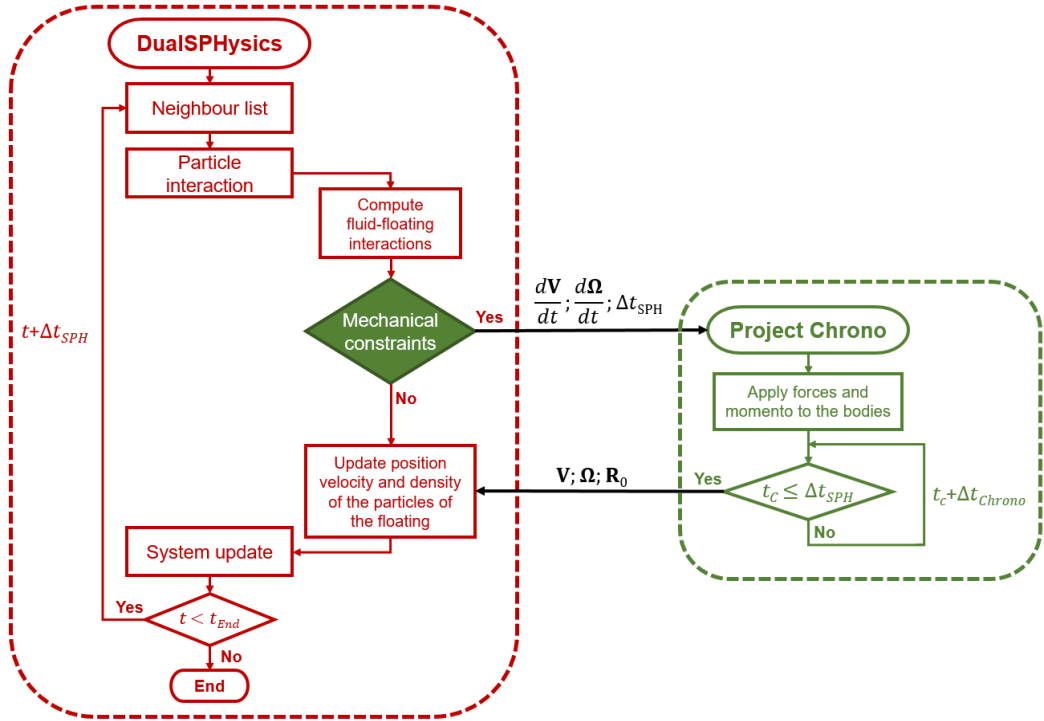

Figure 5. Flow chart of the coupling between DualSPHysics and Project Chrono.

### 4.4. Coupling with MoorDyn

In Section 3.2, the capability of DualSPHysics to simulate fluid-driven objects has been described. The code was shown to be accurate in simulating the interaction of sea waves with floating objects. In real-life problems, floating structures are usually moored to the seabed. Hence, mooring tensions in the floating objects should be included in the SPH solver. The solution is to couple DualSPHysics with a library designed for coupling with other numerical models that already solves mooring dynamics. Therefore, DualSPHysics was coupled with MoorDyn (http://www.matt-hall.ca/moordyn.html), an open-source dynamic mooring line model [88]. MoorDyn discretizes mooring lines as point masses connected by linear spring-damper segments to provide elasticity in the axial direction. It is accurate and efficient, and it includes the contact forces with the seabed.

A complete description of the coupling procedure can be found in Domínguez et al. [89], where the coupling between DualSPHysics and MoorDyn was applied to simulate the motion of a moored floating structure under the action of waves.

Figure 6 shows the flow chart of this coupling approach. It consists in passing the fairlead kinematics to MoorDyn, which then solves the mooring system dynamics. The resulting fairlead tensions are transferred back to DualSPHysics. In this way, MoorDyn calculates the tensions of the mooring lines, which are added as external forces to derive the final resulting motion of the moored floating in the SPH solver.

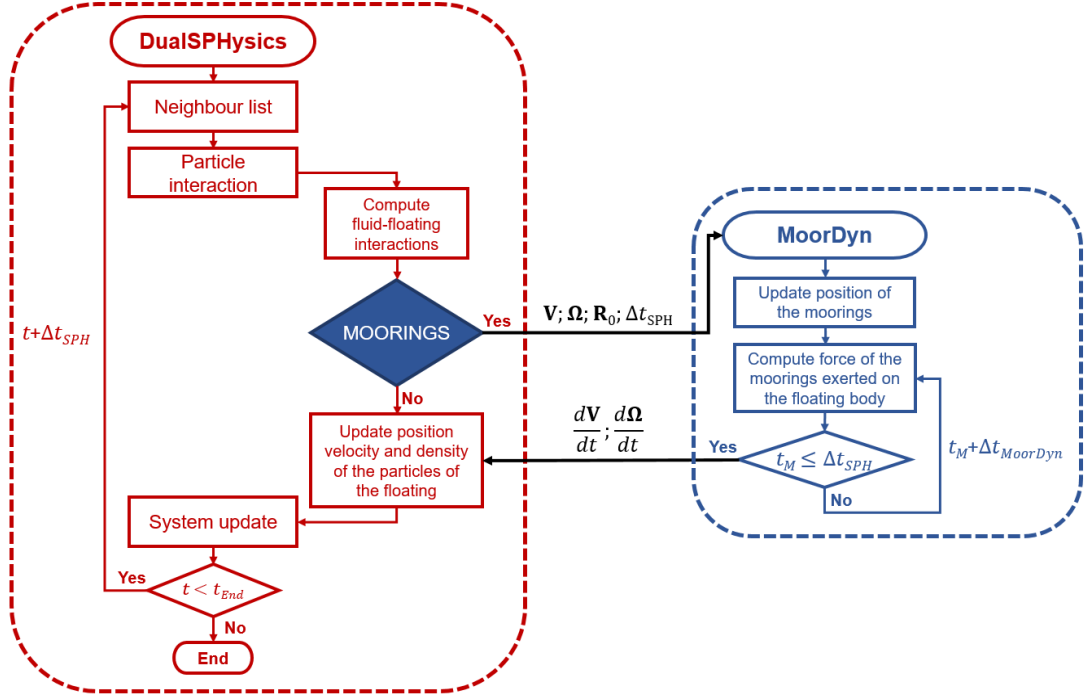

Figure 6. Flow chart of the coupling between DualSPHysics and MoorDyn.

## 5. Code and implementation

DualSPHysics is developed to run on two different hardware architectures: the traditional Central Processing Unit (CPU) and the more recent Graphics Processing Unit (GPU). Therefore, all algorithms and features included in this SPH solver are implemented and optimised for both architectures. The programing language C++ is used to code the SPH formulation presented in Section 3 for CPU execution, but also implements the main program logic and all tasks not related to SPH calculations, such as the initial configuration and input/output data. The GPU execution is performed using the NVIDIA's programming framework Compute Unified Device Architecture (CUDA). Thus, the SPH formulation is also implemented by using the CUDA programming language in kernels or functions executed on GPU rather than CPU.

This dual capability of the DualSPHysics execution provides significant advantages but also some drawbacks. Among the advantages, DualSPHysics can perform large simulations by using the computational power of the latest GPUs by NVIDIA, but can also run simulations on workstations without GPUs by using the current multi-core CPU version of the code. This solver enables fair comparisons of the SPH performance between CPU and GPU, since the same formulation is implemented in an optimised way for both hardware architectures. With the advent of more mature compilers and development environments, the CPU code has become more accessible than the GPU one, especially when modifying and debugging, so it represents an attractive option for users approaching DualSPHysics for the first time. This is paramount since DualSPHysics was developed as an open-source framework for the SPH method and many researchers use it to develop and test new formulations. The main drawback is the higher maintenance cost of the solver and optimisation of new functionalities, since two different implementations must be maintained and different optimisation strategies are needed for each architecture.

## 5.1. Implementation

The SPH implementation in DualSPHysics is centred on three main consecutive tasks: (1) Neighbour List (NL), (2) Particle Interaction (PI), and (3) System Update (SU). These allow the simulation of an interval of physical time (i.e., the simulation step) and their execution is repeated until the total physical time to be simulated is complete, as also shown in Figure 7:

(1) The Neighbour List (NL) prepares the data needed to find the neighbouring particles of each SPH particle efficiently in the next task. Different NL approaches have been considered in the literature [12, 90, 91]. DualSPHysics implements a cell-linked list approach, since this appears to be the most efficient for the simulation of large numbers of particles, as shown in Domínguez et al. [12] for CPU simulations and later in Winkler et al. [91] for GPU simulations. During the NL task, the simulation domain is split up in cubic cells of size $2h$, i.e. the maximum interaction distance between particles. The particles with their data are sorted according to the cell to which they belong and an index array with the first and last particle of each cell is created. The runtime of this task usually represents a small percentage of the total execution time. However, the result of the Neighbour List is critical to maximize the performance on the subsequent task.

(2) The Particle Interaction (PI) is the most time-consuming task. It represents over 90% of the total execution time. During the PI process, the momentum and continuity equations are solved by the interaction of each particle with its neighbouring particles. The neighbour search is performed efficiently since, for each particle, only the neighbouring particles in the adjacent cells will contribute to the particle interaction.

(3) The last task is the System Update (SU), during which the duration of the simulation step is calculated, together with the new values of the properties of SPH particles (position, density, velocity). These properties are computed for the next simulation step according to the values of these properties at the current time step, the results of particle interactions, and the duration of the simulation step.

Other less important but necessary tasks are the initial configuration and data loading, which are executed once at the beginning of the calculation and also during the periodic data output that is usually performed every many simulation steps. More details of the work carried out during each of the above tasks can be found in Crespo et al. [92].

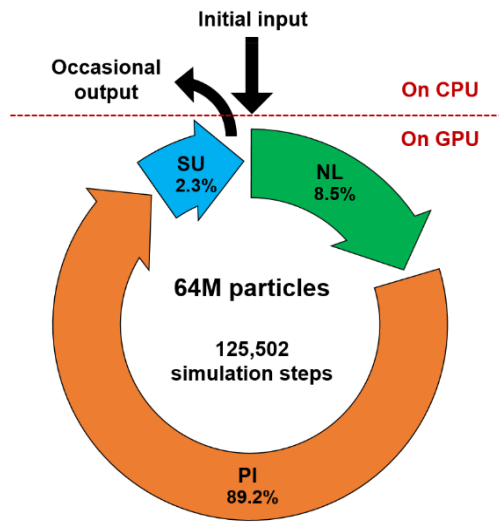

Figure 7. Scheme of the main SPH tasks (NL, PI and SU) during a DualSPHysics execution. The time percentage corresponding to each task is shown during the simulation of a 3-D dam break test case, with 64M particles on a Tesla V100 GPU.

**Programming paradigms:** DualSPHysics is written in C++ following the programming paradigm of Object-oriented programming (OOP), in which the particle data is stored as arrays of structures (AoS) to maximise the performance in parallel executions. The CPU and GPU implementations are available under two different classes (JSphCpu and JSphGpu), both of which inherit from the same class (JSph) with common parts. In addition, the JSphGpu class calls the kernels written in CUDA language. The CPU and GPU codes are very similar to ease the expansion and maintenance of DualSPHysics. However, some differences are inevitable due to the characteristics of each architecture and the different optimisation strategies employed to maximize the performance in each hardware architecture [13]. The CPU implementation uses the application programming interface Open Multi-Processing (OpenMP) to perform parallel execution of SPH method on one or several multi-core CPUs hosted on the same workstation or computation node. Each CPU thread executes in parallel the interactions with the neighbours of a given set of particles, while a CUDA thread is created to compute the interactions of each particle on GPU.

**Memory access:** The CPUs access the main memory of the host directly through different levels of cache memory that minimise the latency of this access. However, the efficient use of the GPU memory's hierarchy is the responsibility of the developer and it is critical to performance. The GPU computation is executed in the device's (GPU) memory and the communication between the main memory and the GPU memory (typically via PCI Express bus) represents the main bottleneck in GPU computing. The GPU implementation of DualSPHysics copies all particle data to the GPU memory at the beginning of the simulation and all calculations are performed on GPU, so that data transfers between the main memory and the device memory are minimized. The particles' data on the GPU memory is only transferred to the CPU memory when this data must be stored in long-term memory, which happens occasionally during the simulation, and thus with negligible impact on performance.

**Precision:** Most of the real variables in DualSPHysics are coded in double precision, but some particle data uses single precision since double precision is generally not necessary and the use of single precision improves the performance, especially during GPU execution. However, the use of double precision is mandatory on position variables for high-resolution simulations with large domains [93]. The precision of particles' position is critical in two phases: the calculation of interaction distance between particles in PI task and the update of particle magnitudes for next simulation step in the SU task. The position variables are implemented in DualSPHysics using double precision and the SU task is performed by using double precision for both architectures (CPU and GPU). However, the relative position of the particle within its cell (using single precision) is computed before the particle interaction and is used to compute the distance between particles on GPU implementation. This optimisation provides a significant performance improvement without loss of accuracy on GPU models where the performance ratio between double and single precision computing is 1/32. This optimisation is not applied to the CPU version of the code since the speedup on CPU is negligible.

## 5.2. Source files and compilation

The DualSPHysics solver consists of more than 240 source files and over 95,000 lines of code. Table 1 shows the source files distributed in four main blocks: i) the DualSPHysics core, with the source files of the SPH formulation and the NL algorithm; ii) the functionalities block containing the code for different DualSPHysics features, such as moored floating bodies, coupling with Chrono, wave generation, etc.; iii) I/O source files to load initial configuration and particle data and to save the simulation results; and iv) general purpose code. All source files are documented following the Doxygen format to facilitate understanding by the SPH scientific community. DualSPHysics is an open source code available in the public repository GitHub (https://github.com/DualSPHysics/DualSPHysics).

The compilation of DualSPHysics code is straightforward since a Makefile for Linux systems and a Visual Studio Project for Windows are included. Also included is a CMakeLists file that allows the automatic compilation of the code using the tool CMake (Cross platform Make). Some pre-compiled libraries are distributed with DualSPHysics. However, they are not critical and a flag compilation allows the user to compile the program without including them. Compiling the DualSPHysics program without CUDA code is also possible in the absence of a GPU.

Table 1. List of source files of DualSPHysics code.

| **DualSPHysics core** | | **DualSPHysics functionalities** | |
|---|---|---|---|
| main.cpp | JSphInOutGridData.h/cpp | *Moored floating bodies:* | *Predefined motion:* |
| DualSphDef.h | JSphInOutPoints.h/cpp | DSphMoorDyn.h | JDsMotion.h/cpp |
| JParticlesDef.h | JSphInOutZone.h/cpp | DSphMoorDynUndef.h | JMotion.h/cpp |
| JPeriodicDef.h | JSphMk.h/cpp | JDsMooredFloatings.h/cpp | JMotionEvent.h |
| JSph.h/cpp | JShifting.h/cpp | JDsFtForcePoints.h/cpp | JMotionList.h/cpp |
| JSphCfgRun.h/cpp | OmpDefs.h | LibDSphMoorDyn.lib/a | JMotionMov.h/cpp |
| JSphInOut.h/cpp | | | JMotionObj.h/cpp |
| | | *Coupling with Chrono:* | JMotionPos.h/cpp |
| **SPH for CPU** | **SPH for GPU** | DSPHChronoLib.h | |
| | | JChronoData.h | *Other features:* |
| FunSphEos.h | FunSphEos_iker.h | JChronoObjects.h/cpp | JDsAccInput.h/cpp |
| FunSphKernel.h | FunSphKernel_iker.h | JChronoObjectsUndef.h | JDsAccInput_ker.h/cu |
| FunSphKernelDef.h | JArraysGpu.h/cpp | dsphchrono.so/lib/dll | JDsDamping.h/cpp |
| FunSphKernelsCfg.h/cpp | JSphGpu.h/cpp | libChronoEngine.so/dll | JDsFixedDt.h/cpp |
| JArraysCpu.h/cpp | JSphGpuSingle.h/cpp | libChronoEngine_parallel.so/dll | JDsGaugeSystem.h/cpp |
| JSphCpu.h/cpp | JSphGpuSingle_InOut.cpp | | JDsGaugeItem.h/cpp |
| JSphCpuSingle.h/cpp | JDebugSphGpu.h/cpp | *Wave generation utils:* | JDsGauge_ker.h/cu |
| JSphCpu_InOut.cpp | JSphGpu_ker.h/cu | JWaveGen.h | JDsInitialize.h/cpp |
| JSphCpuSingle_InOut.cpp | JSphGpu_InOut_iker.h/cu | JWaveGenUndef.h | JDsOutputTime.h/cpp |
| JSphTimersCpu.h | JSphGpuSimple_ker.h/cu | JWaveAwasZsurf.h/cpp | JDsPartsInit.h/cpp |
| JSphBoundCorr.h/cpp | JSphShifting_ker.h/cu | JWaveOrder2_ker.h/cu | JDsPartsOut.h/cpp |
| **NL for CPU** | **NL for GPU** | JWaveSpectrumGpu.h/cpp | JDsphConfig.h/cpp |
| | | JMLPistons.h | JDsPips.h/cpp |
| JCellDivCpu.h/cpp | JCellDivGpu.h/cpp | JMLPistonsGpu.h/cpp | JDsPips_ker.h/cu |
| JCellDivCpuSingle.h/cpp | JCellDivGpuSingle.h/cpp | JRelaxZones.h | JDsSaveDt.h/cpp |
| JCellDivDataCpu.h | JCellDivDataGpu.h | JRelaxZonesGpu.h/cpp | JDsViscoInput.h/cpp |
| JCellSearch_inline.h | JCellDivGpu_ker.h/cu | JRelaxZone_ker.h/cu | JNormalsMarrone.h/cpp |
| | JCellDivGpuSingle_ker.h/cu | LibJWaveGen.lib/a | |
| | JCellSearch_iker.h | | |

| **DualSPHysics I/O** | **General purpose code** | | |
|---|---|---|---|
| *Case configuration:* | Functions.h/cpp | JLog2.h/cpp | JReduSum_ker.h/cu |
| JCaseCtes.h/cpp | FunctionsBasic_iker.h | JMatrix4.h | JSaveCsv2.h/cpp |
| JCaseEParms.h/cpp | FunctionsCuda.h/cpp | JMeanValues.h/cpp | JSimpleNeigs.h/cpp |
| JCaseParts.h/cpp | FunctionsGeo3d.h/cpp | JNumexLib.h | JTimeControl.h/cpp |
| JCasePartsDef.h | FunctionsGeo3d_iker.h | JNumexLibDef.h | JTimer.h |
| JCaseProperties.h/cpp | FunctionsMath.h | JNumexLibUndef.h | JTimerClock.h |
| JCaseUserVars.h/cpp | FunctionsMath_ker.h | LibJNumexLib.lib/a | JTimerCuda.h |
| JCaseVtkOut.h/cpp | JAppInfo.h/cpp | JObject.h/cpp | JVtkLib.h |
| | JBinaryData.h/cpp | JObjectGpu.h/cpp | JVtkLibDef.h |
| *Particle data files:* | JCfgRunBase.h/cpp | JOutputCsv.h/cpp | JVtkLibUndef.h |
| JPartDataBi4.h/cpp | JDataArrays.h/cpp | JPartsLoad4.h/cpp | LibJVtkLib.lib/a |
| JPartDataHead.h/cpp | JException.h/cpp | JRadixSort.h/cpp | JXml.h/cpp |
| JPartFloatBi4.h/cpp | JLinearValue.h/cpp | JRangeFilter.h/cpp | TypesDef.h |
| JPartNormalData.h/cpp | JLinearValueDef.h | JReadDatafile.h/cpp | |
| JPartOutBi4Save.h/cpp | | | |

## 5.3. Performance

With each particle having a constantly changing stencil of neighbouring particles, typically 40 in 2D and around 250 in 3D, the SPH method is computationally demanding. It is therefore essential to apply optimisations such as those discussed above to maximize the performance. This section shows the performance of DualSPHysics according to runtime and Particle Interactions Per Second (PIPS) using the latest sets of optimisation. The runtime of SPH simulations increases rapidly with the number of particles, since each particle interacts with more than 268 neighbouring particles in 3-D (when $2h$ is set to be four times the initial interparticle distance, for example). However, many other factors influence the total execution time. The number of simulation steps per physical seconds grows when the resolution is increased or when there are rapid time-varying events requiring a small timestep during the simulation. The distribution of the fluid in the simulation domain can increase or reduce the total number of interactions between particles. Other important factors affecting the runtime are the physical time to be simulated, the complexity of the chosen formulations and some hardware features. The use of PIPS allows measuring the performance by focusing on the actual number of particle interactions regardless of the number of simulation steps, the amount of particles or their distribution during the simulation.

The DualSPHysics performance was evaluated by simulating the second validation test of SPHERIC (https://spheric-sph.org/tests/test-2). This test case is a dam break flow impacting an obstacle. It was described in Kleefsman et al. [94] and validated with DualSPHysics in Crespo et al. [95]. The first two physical seconds are simulated for different resolutions, where the number of particles ranges from 116,658 to 80,353,962. Simulations were performed using different CPU and GPU models (see Table 2 and 3 respectively). Figure 8 (a) shows the runtime of these simulations after limiting the maximum runtime to 48 hours. Runtimes are compared against the number of particles using different devices, but also the maximum number of particles that each hardware device can simulate in less than 2 days. The fastest GPU is the NVIDIA Tesla V100, which can simulate test cases with 80 million particles in less than 48 hours. The second fastest GPU is the NVIDIA GeForce RTX 2080 Ti, capable of simulating 56 million particles in approximately 36 hours. The fastest CPU can simulate 2.15 million particles in less than 48h.

Figure 8 (b) shows the GPIPS ($10^9$ PIPS) achieved by each device according to the number of particles. This value is about 0.14 using CPU2 (12 execution threads), close to 3 using Kepler GPUs, 12 using the GTX 1080 Ti, 23 using the RTX 2080 Ti and 29 using the Tesla V100. The performance of the computing devices increases according to the size of the problem, since a minimum number of particles is needed to make an efficient and full use of the hardware. This becomes clearer the higher the parallel computing power of the device. We can say that the device is saturated when the performance does not increase as the number of particles increases. The GPIPS value is almost constant when the number of particles is large enough to saturate the calculation device. This saturation value is reached with 100,000 particles on CPU2, 2 million particles on the Kepler GPUs and 12 million particles on the Tesla V100. These results show that the speedup of the Tesla V100 against CPU2 is more than 200x from 4 million particles onwards. The speedup of Tesla V100 against the GeForce RTX 2080 Ti is about 1.3x since the relative position of particles to the cell is used to minimise the use of double precision. This speedup is more than 5x when the particles position in double precision is used, and it is due to a performance ratio double/single precision of 1/32 for the RTX 2080 Ti.

Table 2. Main characteristics of CPU devices.

|  | CPU1: Intel Core i7-6700K | CPU2: Intel Core i7-8700K |
| --- | --- | --- |
| **Code name** | Skylake | Coffe Lake |
| **Technology [nm]** | 14 | 14 |
| **Frequency [GHz]** | 4.0 | 3.7 |
| **Physical cores** | 4 | 6 |
| **Execution threads** | 8 | 12 |

Table 3. Main characteristics of GPU devices.

|  | Tesla K20 | Tesla K40 | GeForce GTX Titan Black | GeForce GTX 1080 Ti | GeForce RTX 2080 Ti | Tesla V100 |
| --- | --- | --- | --- | --- | --- | --- |
| **GPU microarchitecture** | Kepler | Kepler | Kepler | Pascal | Turing | Volta |
| **Global memory [MB]** | 5,063 | 12,207 | 6,084 | 11,264 | 10,989 | 16,128 |
| **CUDA Cores** | 2,496 | 2,880 | 2,880 | 3,584 | 4,352 | 5,120 |
| **Multiprocessors** | 13 | 15 | 15 | 28 | 68 | 80 |
| **CUDA Cores/MP** | 192 | 192 | 192 | 128 | 64 | 64 |
| **Ratio double/single** | 1/3 | 1/3 | 1/3 | 1/32 | 1/32 | 1/2 |
| **GPU Max Clock rate [MHz]** | 706 | 745 | 980 | 1,645 | 1,545 | 1,530 |
| **Memory Clock rate [MHz]** | 2,600 | 3,004 | 3,500 | 5,505 | 7,000 | 877 |
| **Memory Bus Width [bits]** | 320 | 384 | 384 | 352 | 352 | 4,096 |

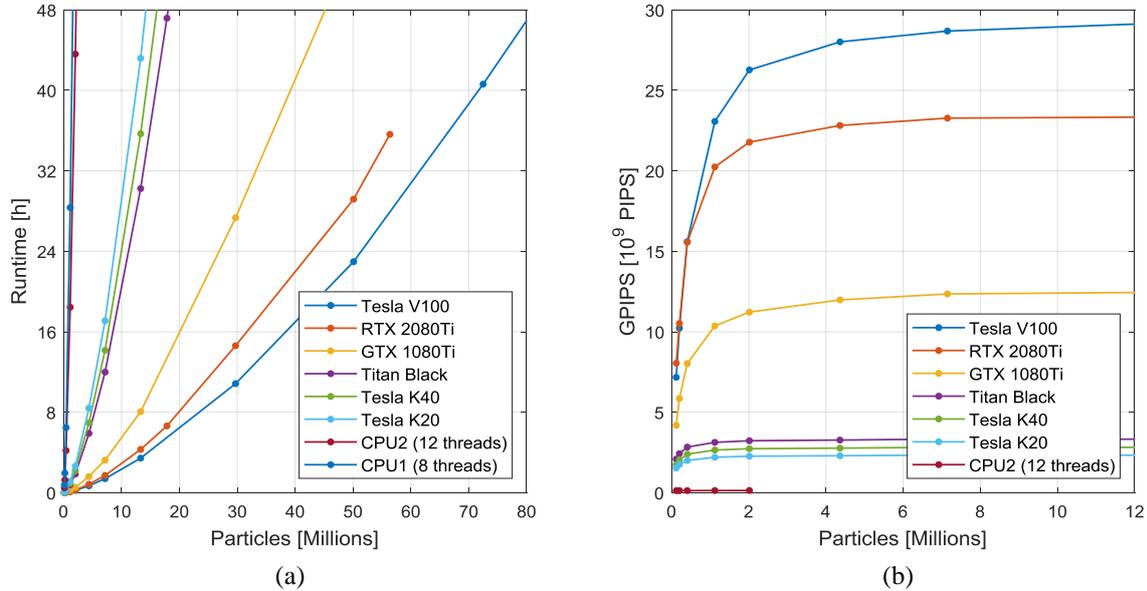

Figure 8. Runtime and Giga Particle Interactions Per Second ($10^9$ PIPS) of the dam break simulation according to number of particles using different hardware devices at (a) and (b) panel respectively.

The performance tests were carried out using cells of size $h$ instead of $2h$ to improve the search for neighbouring particles. This optimisation was introduced in Domínguez et al. [13] and enables CPU and GPU speedups of 1.14x and 1.3x, respectively. However, the memory consumption increases and the maximum number of particles that a GPU can simulate is reduced. The use of cells of size of $2h$ allows simulating 6.05 million particles per Gigabyte of device memory, whereas $h$ allows simulating 5.8 million particles per Gigabyte. Figure 9 shows the maximum number of particles according to memory size of each GPU model. The memory size of GPU devices is a major limitation for simulations requiring large numbers of particles. This limitation is removed in DualSPHysics by using a multi-GPU approach [96, 97]. This approach use Message Passing Interface (MPI) to combine the computation power of different GPUs hosted in the same or different computation nodes. This Multi-GPU version based on DualSPHysics v3.0 allows performing simulations with more than 1,000 million particles with relatively short runtimes [97].

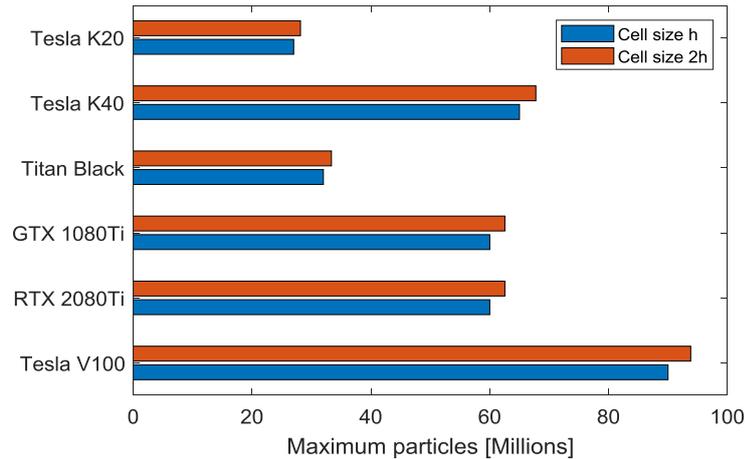

Figure 9. Maximum number of particles according to the memory size of each GPU model.

### 5.4. Pre- and post-processing tools

One of the strengths of the DualSPHysics package is the inclusion of not only the SPH solver, but also pre-processing and post-processing tools. This makes the package standalone and attractive to the user since the entire workflow is provided, including the generation of the initial condition (initial state of the particles in the simulation), the execution of the SPH solver, and the analysis and visualization of the results by computing the physical magnitudes of interest.

The input file (in XML format) includes not only the geometry of the different objects (walls, bottom, wavemaker, floaters, breakwaters, fluid, etc.) but also the definition of constants such as gravity, reference density of water, compressibility coefficients, size of the kernel and initial particle distance. In addition, different options are also selected in the input file, namely, the kernel function, the time integrator, the use of density diffusion term or the choice of the viscosity treatment. Complex geometries can be easily created since a wide variety of commands (labels in the XML file) are available to create different objects: points, lines, triangles, quadrilateral, polygons, pyramids, prisms, boxes, beaches, spheres, ellipsoids, cylinders, etc. The geometry can therefore be defined in the input file. However, the pre-processing tool can also load an external geometry directly from an external file of different formats (VTK, PLY and STL) to create the initial particles.

The pre-processing code is named GenCase and it employs a 3-D Cartesian mesh to deploy the particles. The mesh nodes around the object are defined and then particles are created only in the nodes needed to draw the desired geometry. The geometry of the case is defined independently from the inter-particle distance. This allows the discretization of each test case with a different number of particles simply by varying the resolution (or initial particle distance).

Once the simulation is finished, the package provides different post-processing tools with a twofold aim: visualisation and computation of magnitudes of interest. The output files of the SPH solver contain the state of the particles (position, velocity and density/pressure) at different time instants. The visualisation tools allow the representation of particles in VTK format, where the velocity and density/pressure fields can be shown. Additionally, different post-processing tools are available in the package to compute physical quantities including: i) free-surface elevation at wave gauges, ii) velocity and pressure at numerical gauges, iii) vorticity in the fluid domain, iv) force exerted by the fluid onto a boundary (fixed or floating structure), v) motions and rotations of fluid-driven objects, and vi) inflow and outflow rates through domains defined by the user.

The range of applications of SPH-based simulations is growing for industrial problems. The use of a graphical user interface (GUI) is key to encourage new users to work with SPH codes. Therefore, a GUI named DesignSPHysics and shown in Figure 10 has been developed, which integrates the whole workflow described before: pre-processing, SPH solver and post-processing. DesignSPHysics is developed as open-source software. FreeCAD is chosen as the host 3-D modelling software for the plug-in using Python as the default scripting language and QT as its GUI framework. DesignSPHysics is presented as part of the DualSPHysics package, meaning that any user already familiar with the command line interface and XML-driven case generation can easily understand how the GUI works. However, the main objective of DesignSPHysics is that new users, not familiar with DualSPHysics, can create new cases from scratch and run simulations.

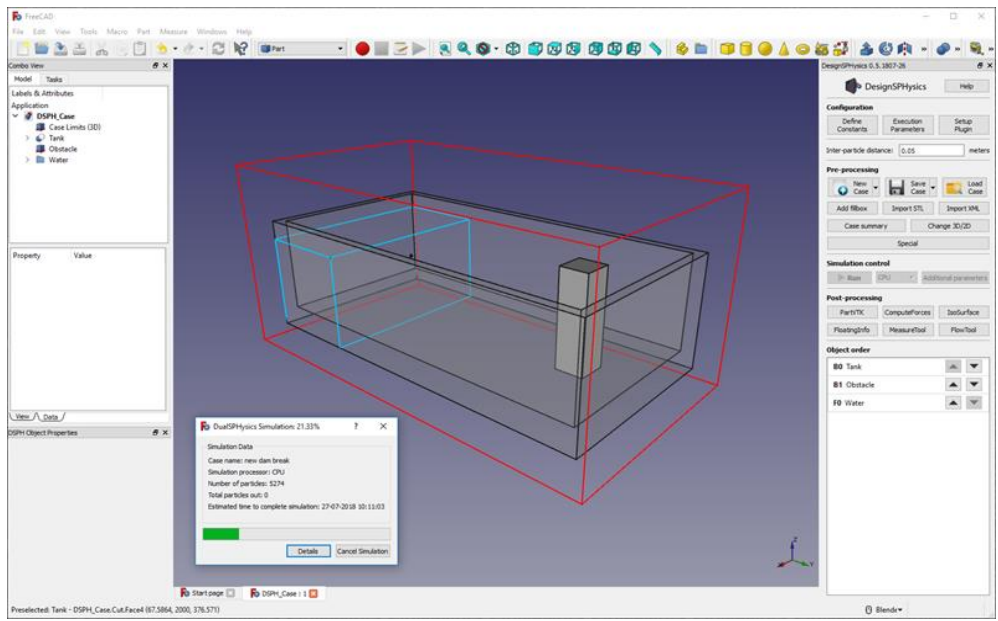

Figure 10. Graphical User Interface named DesignSPHysics.

# 6. Applications

The features developed in the different versions of DualSPHysics have been validated with analytical solutions, numerical benchmarks and experimental results demonstrating accuracy and convergence as it will be described in the following examples retrieved from the literature. Table 4 includes some relevant works focused on DualSPHysics published over the last few years, with special emphasis on validated applications.

Table 4. Examples in the literature over the last years using DualSPHysics.

| Manuscript | Main topic | Main achievements |
|---|---|---|
| Altomare et al., 2015a [78] | Coastal structures | Regular and random waves colliding with coastal structures<br>Comparison with experimental data and analytical solutions |
| Altomare et al., 2015b [98] | Coastal structures | Coupling with SWASH (multi-layer piston)<br>Comparison with experimental data |
| Pringgana et al., 2016 [99] | Coastal structures | Tsunami-induce bore impact on onshore structure<br>Comparison with semi-empirical approaches |
| Altomare et al., 2017 [42] | Coastal structures | Automatic wave generation<br>Passive and active wave absorption<br>Comparison with experimental data and analytical solutions |
| Altomare et al., 2018 [81] | Coastal structures | Coupling with SWASH (relaxation zone)<br>Comparison with experimental data and theoretical solutions |

| González-Cao et al., 2018a [100] | Coastal structures | Comparison with experimental data and mesh-based models |
|---|---|---|
| Zhang et al., 2018 [101] | Coastal structures | Regular waves interacting with porous breakwater<br>Time series of wave run-up<br>Comparison with experimental data |
| Rota-Roselli et al., 2018 [102] | Coastal structures | Optimization of model setup for wave propagation through genetic algorithms<br>Comparison with theoretical solutions |
| Verbrugghe et al., 2019b [82] | Coastal structures | Coupling with OceanWave3D<br>Open boundary conditions |
| Domínguez et al., 2019a [43] | Coastal structures | Solitary waves (different theories)<br>Comparison with experimental data and analytical solutions |
| Lowe et al., 2019 [103] | Coastal structures | Spilling and plunging wave breaking<br>Comparison with experimental data |
| Subramaniam et al., 2019 [104] | Coastal structures | 3D modeling of wave-run on convex and concave dikes<br>Comparison with experimental results and mesh-based methods |
| Canelas et al., 2015 [105] | Fluid-driven objects | Buoyancy of floating bodies<br>Comparison with experimental data and mesh-based models |
| Canelas et al., 2016 [83] | Fluid-driven objects | DEM technique<br>Interaction between floating bodies<br>Comparison with experimental data |
| Canelas et al., 2017 [106] | Fluid-driven objects | DEM technique for Debris flow<br>Interaction between floating bodies<br>Comparison with experimental data |
| Domínguez et al., 2019b [89] | Fluid-driven objects | Simulation of freely and moored floating box<br>Coupling with MoorDyn<br>Comparison with experimental data |
| Canelas et al., 2018 [87] | Multiphysics problems | Coupling with Chrono<br>Comparison with experimental data |
| Crespo et al., 2017 [107] | Wave Energy Converters (WECs) | Simulation of fixed and floating oscillating water column with open chamber<br>Comparison with experimental data |
| Verbrugghe et al., 2018 [80] | Wave Energy Converters (WECs) | Coupling with OceanWave3D<br>Simulation of oscillating water column and point absorber |
| Brito et al., 2020 [108] | Wave Energy Converters (WECs) | Simulation of oscillating wave surge converter<br>Coupling with Chrono<br>Comparison with experimental data |
| González-Cao et al., 2018b [109] | Hydraulic problems | Safety of spillways<br>Open boundary conditions |
| Novak et al., 2019 [110] | Hydraulic problems | Vertical slot fishway<br>Comparison with experimental data |
| Mokos et al., 2015 [65] | Multi-phase air-water | GPU implementation of gas-liquid solver<br>Comparison with experimental data |
| Mokos et al., 2017 [63] | Multi-phase air-water | Gas-liquid solver with shifting algorithms<br>Comparison with experimental data |
| Fourtakas and Rogers, 2016 [73] | Multi-phase soil-water | State-of-the-art non-Newtonian and multi-phase solver<br>Comparison with experimental data |
| Zubeldia et al., 2018 [111] | Multi-phase soil-water | Improvements over the erosion characteristics of the multi-phase soil-water solver<br>Comparison with experimental data |

One of the major fields of application of DualSPHysics has been coastal engineering, where the model has been employed as a complementary tool to experimental measurements. A fundamental requirement for proper capturing of wave-structure interaction phenomena is the accurate modelling of wave generation and propagation. Altomare et al. [42] implemented detailed and accurate wave generation and absorption schemes for long-crested second-order waves (Section 3.1). The validation of the results from the numerical model was carried out against analytical solutions and experimental results for both regular and random waves. Zhang et al. [101] studied wave run-up for the design of coastal defences. First, run-up time series were validated against experiments using a porous breakwater for which the exact geometry was used in the numerical tank (Figure 11). Then, the model was applied to analyse a real sea dike from the coast of Chongwu in China.

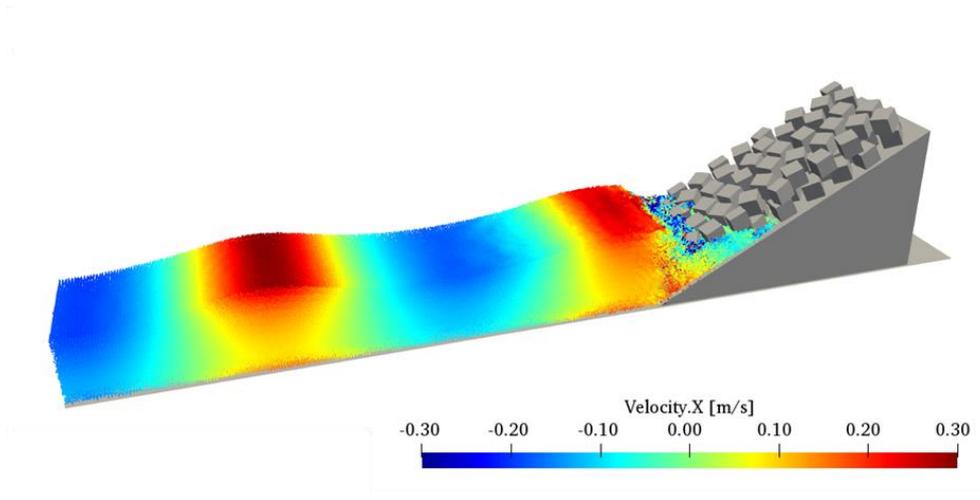

Figure 11. Simulation snapshot of an armour breakwater presented in Zhang et al. [101]. Colour of the particles represent the longitudinal velocity. Complete simulation here: https://youtu.be/W9uIsGomTvE

As previously described in Section 4.1, several efficient wave propagation models can be coupled with SPH-based models to reduce the computational cost in order to propagate the incoming wave far enough from the coastal structure. The two-way coupling with OceanWave3D and based on open boundary conditions was described in Verbrugghe et al. [82], where the experimental data from Ren et al. [112] was used to validate the method and to study the response of a floating box in terms of its three degrees of freedom (heave, surge and pitch).

The accuracy of DualSPHysics to simulate fluid-driven objects, including collisions between solids, was analysed in Canelas et al. [83]. The solid-solid interactions solved with Discrete Element Method (DEM) in conjunction with DualSPHysics was first introduced in that work. Free surface flows and rigid body dynamics were solved in the same meshless framework. An experimental campaign was used to validate this coupling approach. The experiments represented a dam break impacting a set of blocks disposed in different configurations. Blocks were tracked and experimental positions were compared with the numerical ones. The model showed to be accurate and capable of handling highly complex interactions. Later, this SPH-DCDEM approach was tested by comparing numerical results with an experiment designed for stony debris flows in a slit check dam in Canelas et al. [106]. The simulations (Figure 12) aim to mimic the experiment where the solid material is introduced through a hopper, assuring a constant solid discharge. DCDEM was used to solve thousands of time-evolving interactions due to the many simultaneous contacts between the particles that form the different stones. Validation included the comparison of the sediment trapping efficiency, which is obtained by measuring the solid discharges at a position sufficiently upstream and immediately downstream of the dam. Numerical results showed a good agreement with experimental data.

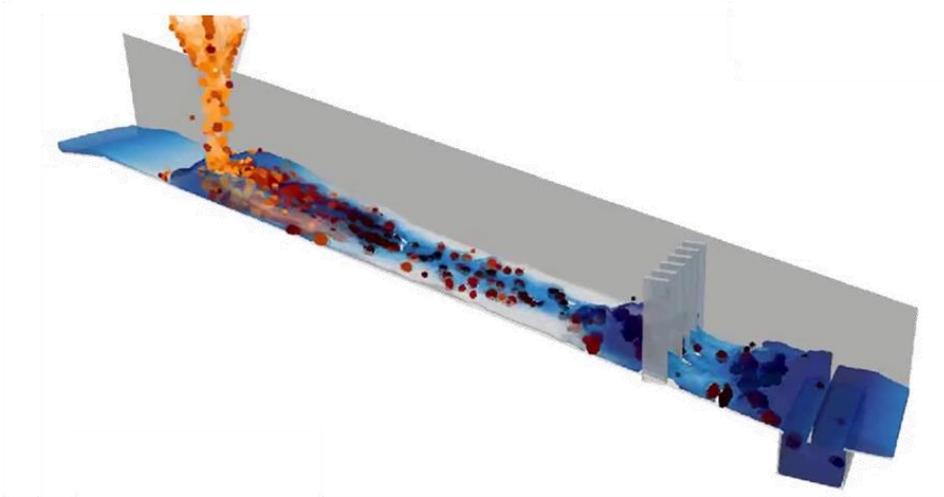

Figure 12. Snapshot of the Simulation of debris flow simulation performed in Canelas et al. [106]. Complete simulation here: https://youtu.be/trbkKUTuB9I.

The validation of moored floating objects under the action of waves using DualSPHysics can be found in Domínguez et al. [89], where the numerical results of nonlinear waves interacting with a freely floating box were compared with the experimental data of Ren et al. [112]. Good agreement was obtained for the motions of the box (heave, surge and pitch). The coupling with the MoorDyn library was validated against data from scale model tests generated during the experimental campaigns for the European MaRINET2 EsflOWC project. The physical tests consisted of a floating box moored to the bottom through four chains. Figure 13 includes an instant of the simulation of the floating box moored with four lines. Numerical and experimental motions of the box and tensions in the lines were in good agreement for different regular waves. However, the model still needs to be validated under several sea states (also irregular waves) including those where an incoming wave train approaches the natural frequency of the structure.

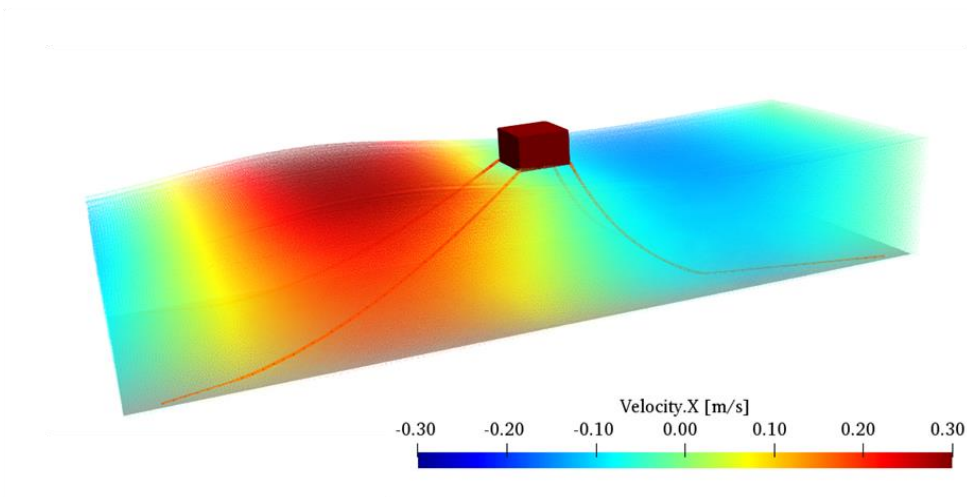

Figure 13. Simulation snapshot of the moored box under regular waves studied in Domínguez et al. [89]. Colour of the particles represent the horizontal velocity. Complete simulation here: https://youtu.be/UaRQII8UgQs.

A detailed description of the coupling between DualSPHysics and Project Chrono was included in Canelas et al. [87]. A dam break flow impacting a supported platform was simulated in order to validate frictional contacts. The numerical times for movement initiation and the angle inversion times were consistent with the experimental ones. Secondly, different pendulum systems were tested, including a system set-up with a hinge (gravity pendulum) and another one with a spring (spring pendulum). The numerical angular rotation was in agreement with the experimental results. Canelas et al. [87] also included some working examples and interesting applications of this coupling with Project Chrono. Figure 14 includes a snapshot of the simulation of the WaveStar machine, a wave energy converter (WEC) that consists of a row of half-submerged buoys. The oscillatory motion of the buoys interacting with waves is converted to electricity with a Power Take-Off system (PTO) that uses the rotational speed of the arm connected to the buoy. DualSPHysics coupled with Project Chrono can solve both the interaction of the incoming waves with the buoys and the hinge restrictions in the connections between the arms and the buoys.

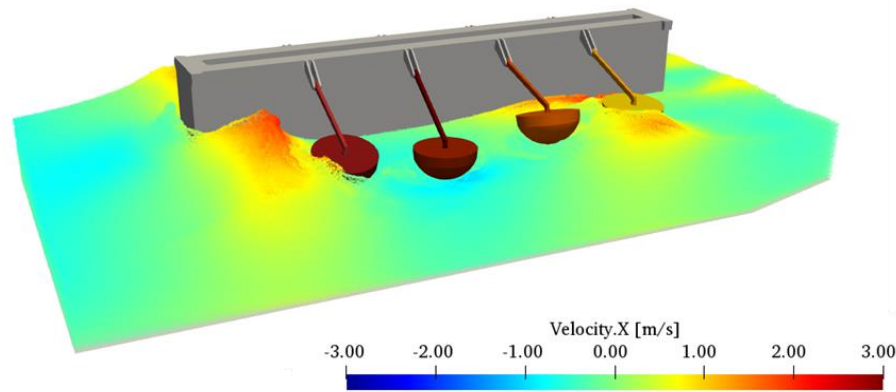

Figure 14. A snapshot of the simulation of the WaveStar machine presented in Canelas et al. [87]. Colour of the particles represent the horizontal velocity. Complete simulation here: https://youtu.be/JeD89PMiLLQ.

Latest versions of DualSPHysics that include coupling with MoorDyn and Project Chrono allow studying the hydrodynamic response of different wave energy converters (WEC), including the numerical behaviour of their PTO. One of the examples that include an exhaustive comparison with experimental results can be found in Brito et al. [108], where an oscillating wave surge converter (OWSC) was studied. The version of DualSPHysics employed in Brito et al. [108] resolved the wave-flap interaction with SPH and the flap-mechanical constraints interaction with Project Chrono. The code was validated using the experimental data of an OWSC with mechanical constraints, including the effects of nonlinear constraints of the hydraulic PTO system and the frictional contacts between flap and bearings. The comparison between numerical and experimental results (including free-surface elevation and angular velocity of the flap) proved that the numerical simulation can predict the hydrodynamic response of the OWSC accurately for both unidirectional regular and irregular waves. Figure 15 depicts a frame of the simulation of OWSC using DualSPHysics. Once validated, the numerical simulation tool was applied to study the influence of several mechanical constraints, PTO damping characteristics and flap inertia on the efficiency of the device.

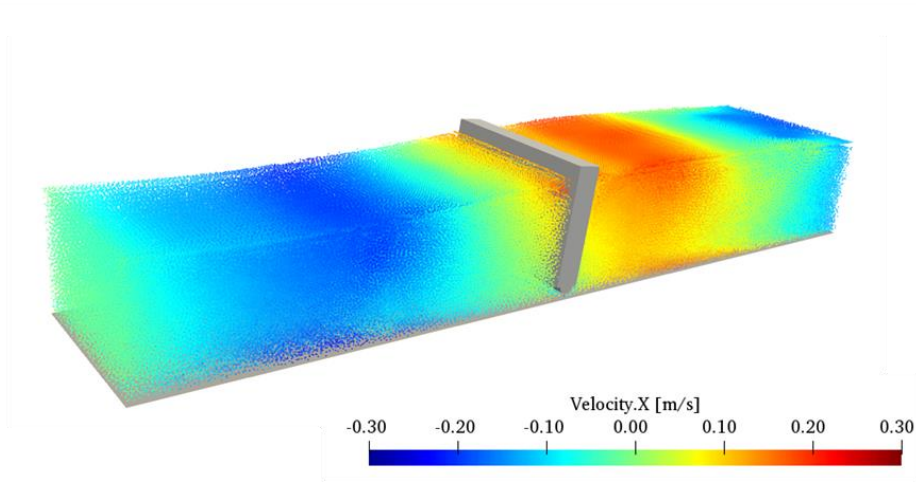

Figure 15. Simulation snapshot of the OWSC analysed in detail in Brito et al. [108]. Colour of the particles represent the horizontal velocity. Complete simulation here: https://youtu.be/-syzUy8qVTE.

Hydraulic engineering is another interesting and common field of application of DualSPHysics. A numerical study of a vertical slot fishway (VSF) in 3-D was performed in Novak et al. [110]. This work used the novel open boundary conditions based on buffer regions to reproduce the free surface turbulent subcritical flow presents in this VSF application. The results show a good agreement with experimental data in discharges, velocity profiles and water elevation at different locations. The Figure 16 shows a snapshot of the VSF simulation where the colour of particles represents the longitudinal velocity.

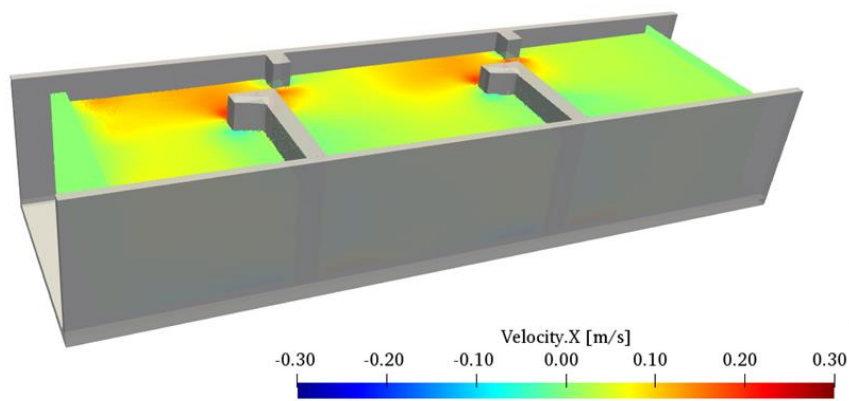

Figure 16. Snapshot of the simulation of the Vertical Slot Fishway analysed in Novak et al. [110]. Colour of the particles represent the horizontal velocity.

Many applications cannot be adequately simulated by a single-phase representation. Since version 3.4 of DualSPHysics, the multi-phase capability has been added to the code to respond to a growing number of requests for this functionality from users, but also to extend the applicability of the code to liquid-gas and liquid-sediment interactions problems. The code has been used for the simulation of 3-D dam breaks impacting structures and fuel-tank sloshing [63, 65], where the presence of the gas phase is necessary to obtain a closer agreement with experimental data. During this process, the hardware acceleration and increased resolution led to the formation of unphysical voids in the flow, which required the development of a novel multi-phase shifting algorithm [63]. Figure 17 shows the results with the application of multi-phase shifting. The non-Newtonian formulation presented in this paper in Section 3.5 is the result of multiple developments for liquid-sediment interaction where water flows over erodible sediment [73]. This requires a combination of (1) unyielded sediment, (2) a yield surface predicted using the Drucker-Prager model, (3) yielded sediment predicted using the Herschel-Bulckley-Papanastiou (HBP) non-Newtonian flow model, (4) a bed load predicted by the Sheilds' criterion, (5) fluid with entrained sediment [111]. Figure 18 shows the results from Fourtakas and Rogers [73], simulating a dam break simulation capturing 3-D experimental erodible sediment profiles with consistent repeatability. The results showed encouraging agreement with experiments for the time history of the water level and the final sediment profiles along the tank.

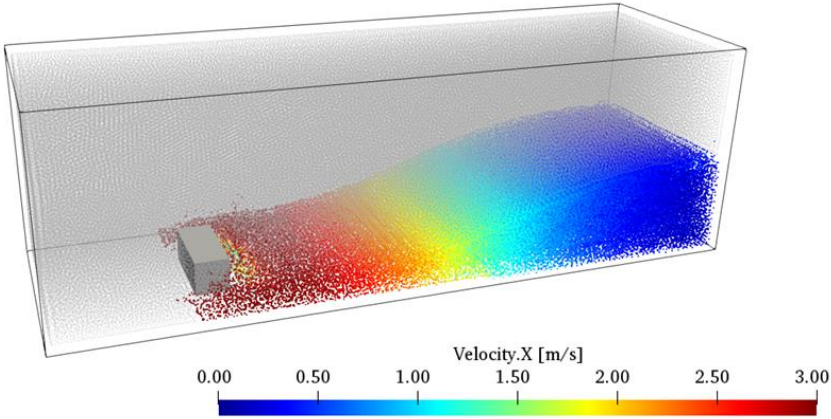

Figure 17. Snapshot of the simulation of the dam break impacting an obstacle using the multi-phase air-water code presented in Mokos et al. [63]. Colour of the particles represent the horizontal velocity. Complete simulation here: https://youtu.be/Q2RZWo-APC8.

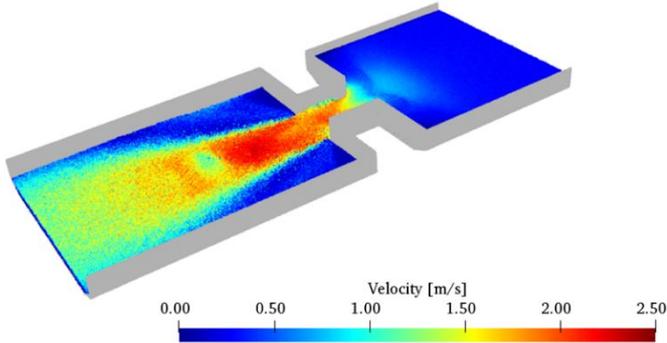

Figure 18. Simulation snapshot of a 3-D dam break over an erodible bed presented in Fourtakas and Rogers [73]. Colour of the particles represent the velocity.

As mentioned above, DualSPHysics is an open source code that can be adapted by users to solve particular engineering problems that were not in the initial scope of the solver. Table 5 shows some examples of these tailored solutions that were implemented with or without the collaboration of the developers. There is a wide range of applications that goes from fundamental problems that cover the development of an incompressible version of the code [113], new boundary conditions [114] or the use of adaptive spatial resolution [115]; to applications like heat transfer [116] or the decay of radionuclides [117]. Due to the different nature of the phenomena under study these new developments do not necessarily appear in further releases.

Table 5. Tailored solutions based on DualSPHysics.

| Manuscript | Main topic | Main achievements |
| --- | --- | --- |
| Mayoral-Villa et al., 2016 [117] | Radionuclides in aqueous dissolution | SPH technique to model advection, diffusion, and radioactive decay<br>Benchmark test cases |
| Chow et al., 2018 [113] | Incompressible SPH | Implementation of a strictly incompressible SPH solver<br>Comparison with experimental data |
| Green et al., 2018 [114] | Boundary conditions | Ghost particles with correction of Adami et al [118]<br>Long duration sloshing simulations<br>Comparison with experimental data |
| Hosain et al., 2019 [116] | Heat Transfer modelling | Benchmarking of thermal solutions, comparison with Finite Volume Method<br>Industrial processes |
| Leonardi et al., 2019 [115] | Variable resolution | Variable particle resolution approach with explicit kernel correction |

# 7. Outlook

This paper has presented the latest novel developments of the community open-source SPH code DualSPHysics. The paper has outlined the SPH formulation, code structure and validations of the latest version 5.0. The sketch presented in Figure 19 summarizes the main functionalities of the DualSPHysics package and the present range of applications. Many of the novel developments presented are motivated towards solving the SPHERIC Grand Challenges: GC2-Boundary conditions, GC4-Coupling, GC5-Applicability to industry. Indeed, GC5 perhaps receives less attention than it deserves, and therefore it needs special efforts to encourage the uptake and use of SPH as a simulation tool in industry.

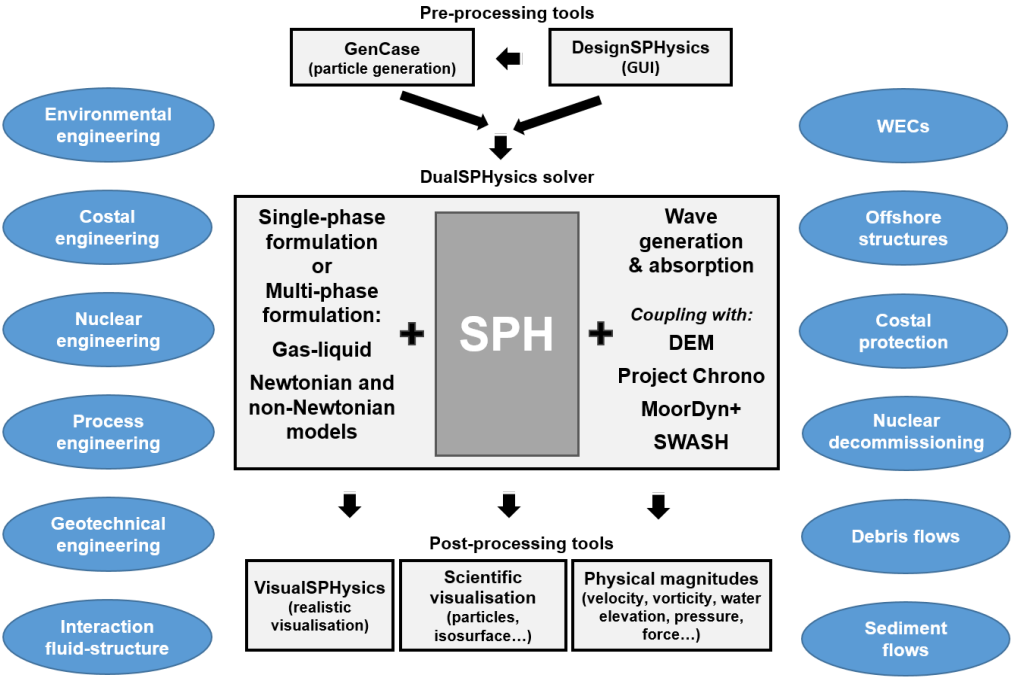

Figure 19. Sketch of DualSPHysics package with components, functionalities and applications.

The resolution required to capture the correct physics of engineering simulations is often beyond the computational capabilities of GPU-accelerated or massively parallel CFD software packages. It is potentially more challenging for SPH since its Lagrangian nature means that the physical processes that have the potential to be implemented offer a tantalising prospect but come with resolution and hence computational demands that are beyond most software packages. Examples include multi-phase applications where the physics can be extremely challenging over a large range of time and length scales. The development of DualSPHysics is specifically aimed at meeting these challenges and will continue to develop in this direction.

Within the context of the rapidly advances and diversification in the field of computer hardware, DualSPHysics is highly optimised to exploit the Nvidia GPU hardware with optimised CUDA kernels, templates, and other optimisations. When general-purpose computing on graphics processing units (GPGPU) first gained traction in the scientific computing community, GPUs were entirely separate devices from the CPU, with CUDA kernels being launched solely from the CPU. Now, the general trend in the development of hardware for computing is for increased integration of the streaming multiprocessors within the motherboard. The most recent multi-GPU computer released by Nvidia (A100) features 16 GPUs on a single board connected with a NVLink (https://www.nvidia.com/en-us/data-center/nvlink).  Similar developments are taking place with the other major hardware vendors. It is clear that increasing integration will be the future of scientific computing with GPGPUs. In order to exploit this, DualSPHysics will evolve to meet the emerging challenges of new hardware. Ultimately, it is possible that SPH will offer a viable method for the biggest challenge that the scientific computing community is now facing: developing algorithms to exploit the ever-emerging hardware for the newborn Exascale computing revolution.

## Acknowledgements

This work was partially financed by the Ministry of Economy and Competitiveness of the Government of Spain under project "WELCOME ENE2016-75074-C2-1-R" and financed by Xunta de Galicia (Spain) under project ED431C 2017/64 "Programa de Consolidación e Estructuración de Unidades de Investigación Competitivas (Grupos de Referencia Competitiva)" co-funded by European Regional Development Fund (ERDF). We are grateful for funding from the European Union Horizon 2020 programme under the ENERXICO Project, Grant Agreement No. 828947 and the Mexican CONACYT- SENER Hidrocarburos Grant Agreement No. B-S-69926.

Dr. J. M. Domínguez acknowledges funding from Spanish government under the program "Juan de la Cierva-incorporación 2017" (IJCI-2017-32592). Dr. C. Altomare acknowledges funding from the European Union's Horizon 2020 research and innovation programme under the Marie Sklodowska-Curie grant agreement No.: 792370.

**Conflict of Interest:** The authors declare that they have no conflict of interest.